\documentclass[3p]{elsarticle}
\usepackage{amssymb}
\usepackage{amsmath}
\usepackage{amsthm}
\usepackage{graphicx}
\usepackage{mathrsfs}
\journal{Journal of Computational Physics}
\theoremstyle{definition} \newtheorem{example}{Example}
\theoremstyle{definition} 
\theoremstyle{plain}

\renewcommand{\dj}{d\kern-0.4em\char"16\kern-0.1em}
\renewcommand{\DJ}{\raise0.3ex\hbox{-}\kern-0.36em D}
\begin{document}
\begin{frontmatter}

\title{Second-order accurate finite volume method for well-driven flows}

\author[cerni]{M. Dotli\'{c}\corref{cor1}}
\ead{milan.dotlic@jcerni.co.rs}
\author[cerni]{D. Vidovi\'{c}}
\ead{dragan.vidovic@jcerni.co.rs}
\author[cerni]{B. Pokorni}
\ead{boris.pokorni@jcerni.co.rs}
\author[rudarski]{M. Pu\v si\'c}
\ead{milenko.pusic@jcerni.co.rs}
\author[cerni]{M. Dimki\'c}
\ead{jdjcerni@jcerni.co.rs}

\address[cerni]{Jaroslav \v Cerni Institute, Jaroslava \v Cernog 80, 11226 Pinosava, Belgrade, Serbia}
\address[rudarski]{University of Belgrade, Faculty of Mining and Geology, \DJ u\v sina 7, 11000 Belgrade, Serbia}
\cortext[cor1]{Corresponding author. Tel. +381 64 274 5246; fax: +381 11 390 6480}

\begin{abstract}
We consider a finite volume method for a well-driven fluid flow in a porous medium.
Due to the singularity of the well, modeling in the near-well region with standard numerical schemes results in a completely wrong total well flux and an inaccurate hydraulic head. 
Local grid refinement can help, but it comes at computational cost.
In this article we propose two methods to address the well singularity.
In the first method the flux through well faces is corrected using a logarithmic function, in a way related to the Peaceman model.
Coupling this correction with a non-linear second-order accurate two-point scheme gives a greatly improved total well flux, but the resulting scheme is still inconsistent.
In the second method fluxes in the near-well region are corrected by representing the hydraulic head as a sum of a logarithmic and a linear function.
This scheme is second-order accurate.
\end{abstract}
\end{frontmatter}
\section{Introduction}
\label{intro}
The stationary groundwater flow equation is obtained by substituting the Darcy law
\begin{equation}
{\bf u}=-\mathbb{K}\nabla h \quad \text{in} \quad \Omega\label{darcy}
\end{equation}
into the continuity equation
\begin{equation}
\nabla\cdot{\bf u}=g_{\text{s}}\label{massCons},
\end{equation}
where ${\bf u}$ is the Darcy velocity, $g_{\text{s}}$ describes sources and sinks, $\mathbb{K}$ is the hydraulic conductivity tensor, $h$ is the hydraulic head, and $\Omega \subset \mathbb{R}^{3}$ is a bounded domain. In this paper we assume that the hydraulic conductivity is isotropic, so that $\mathbb{K}=KI$.

We consider the following boundary conditions:
\begin{equation}
h=g_{\text{D}} \quad \text{on} \quad \Gamma_{\text{D}},
\end{equation}
\begin{equation}
{\bf u}\cdot{\bf n}=g_{\text{N}} \quad \text{on} \quad \Gamma_{\text{N}},\label{Neumann}
\end{equation}
where $\partial \Omega = \Gamma_{\text{D}}\cup \Gamma_{\text{N}}$ is the domain boundary,  $\Gamma_{\text{D}}\cap\Gamma_{\text{N}}=\emptyset$, $\Gamma_{\text{D}}\neq \emptyset$, $\Gamma_{\text{D}}=\bar{\Gamma}_{\text{D}}$ and ${\bf n}$ is a unit vector normal to $\partial \Omega$ pointing outwards.  

\begin{figure}
\begin{center}
  \includegraphics[width=63mm]{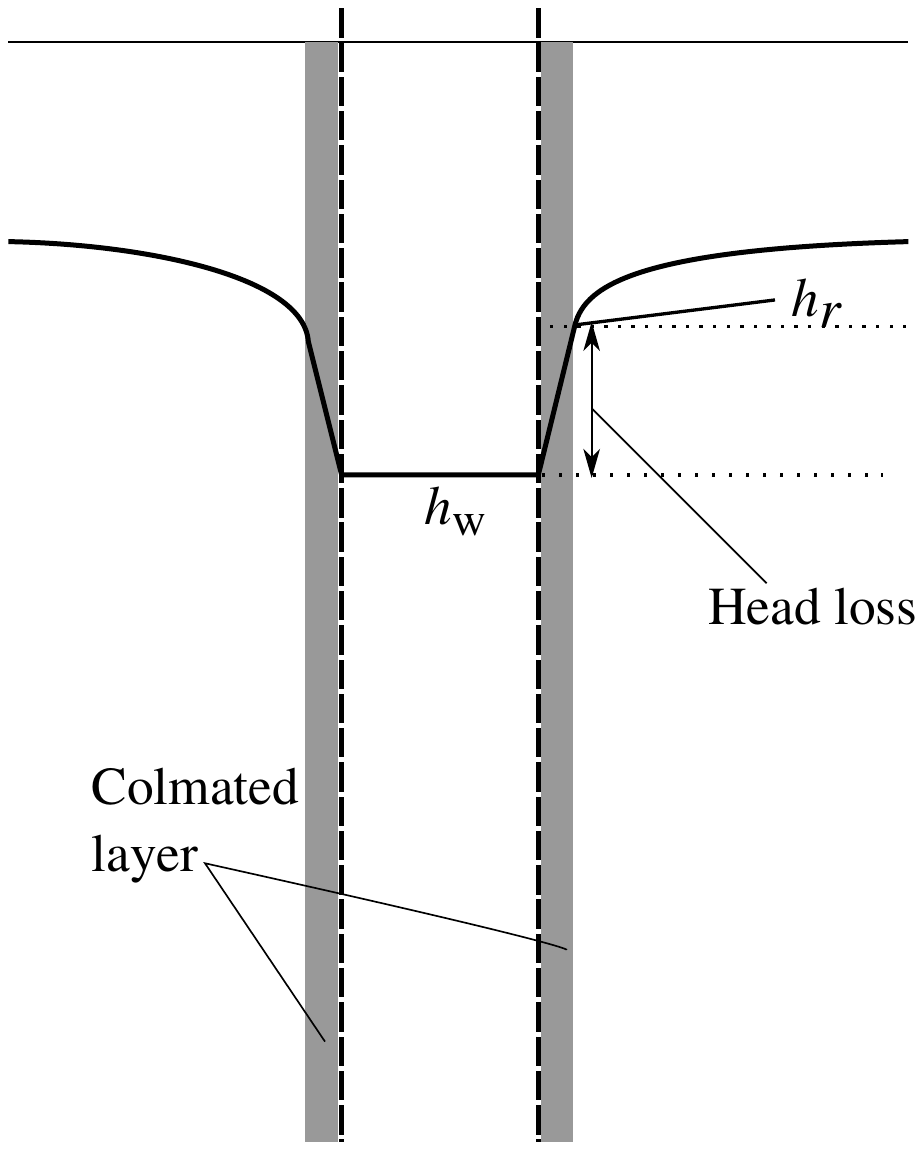}
\caption{Head loss due to colmated layer.}
\label{skinEffect}       
\end{center}
\end{figure}

A colmated layer, also known as the skin effect, is formed along well walls due to well clogging \cite{Dim11a,Dim14}. This causes an additional hydraulic resistance (see Fig. \ref{skinEffect}). As a result, the flux density through the well filter is
\begin{equation}
u=\Psi (h_{r}-h_{\text{w}}),\label{colmation}
\end{equation}
where: $h_{\text{w}}$ is the hydraulic head inside the well, $h_{r}$ is the hydraulic head just outside the colmated layer (see Fig. \ref{skinEffect}), $r$ is the well radius, and $\Psi=K_{\text{c}}/d_{\text{c}}$ is the transfer coefficient, while $K_{\text{c}}$ and $d_{\text{c}}$ are the unknown conductivity and thickness of the colmated layer, respectively. The physical colmated layer thickness is assumed to be small, so that this layer can be modelled as an infinitely thin film of finite $\Psi$. 

Hydraulic head varies logarithmically and its gradient changes sharply in the well vicinity (Fig. \ref{skinEffect}). Thus, linear approximation of hydraulic head is inappropriate on coarse grids and numerical methods based on it are inaccurate in the near-well region.

Accurate modeling in the near-well region is important in reservoir engineering. Flow in the entire reservoir is induced mainly by wells, therefore poor near-well modeling results in accuracy loss throughout the model.

Numerous families of second-order accurate numerical methods are applicable to porous media flows. Here we consider non-linear two-point approximations \cite{Dan09,Lep05,Lip07,Vas08,Vid11,Vid13,Vid14,Yua08}. Although there is no proof that these methods are second-order accurate \cite{Dro14}, numerical tests show second-order accuracy for the hydraulic head and first-order accuracy for the fluxes. These schemes preserve positivity of the solution, but at the price of having to solve a non-linear system even when the problem is linear. 
Nevertheless, linear approximation is deployed and therefore the accuracy is lost on coarse grids if a well is present.

Local grid refinement can alleviate the problem \cite{Mun10}. However, this comes at a computational cost.

Methods for well modeling have been widely discussed in the literature \cite{Che09,Din01,Din04,Dur00,Pea78,Pea83,Mun10}. A commonly used method is the Peaceman model \cite{Din01,Pea78,Pea83}. This approach was originally formulated for finite differences, with a well placed in a cell center. It has been extended to various other discretization methods \cite{Che09}. Peaceman model introduces an additional equation which yields a greatly improved flow rate, but it does not improve the accuracy of the hydraulic head around the well. 

\begin{figure}[htb]
\begin{center}
  \includegraphics[width=110mm]{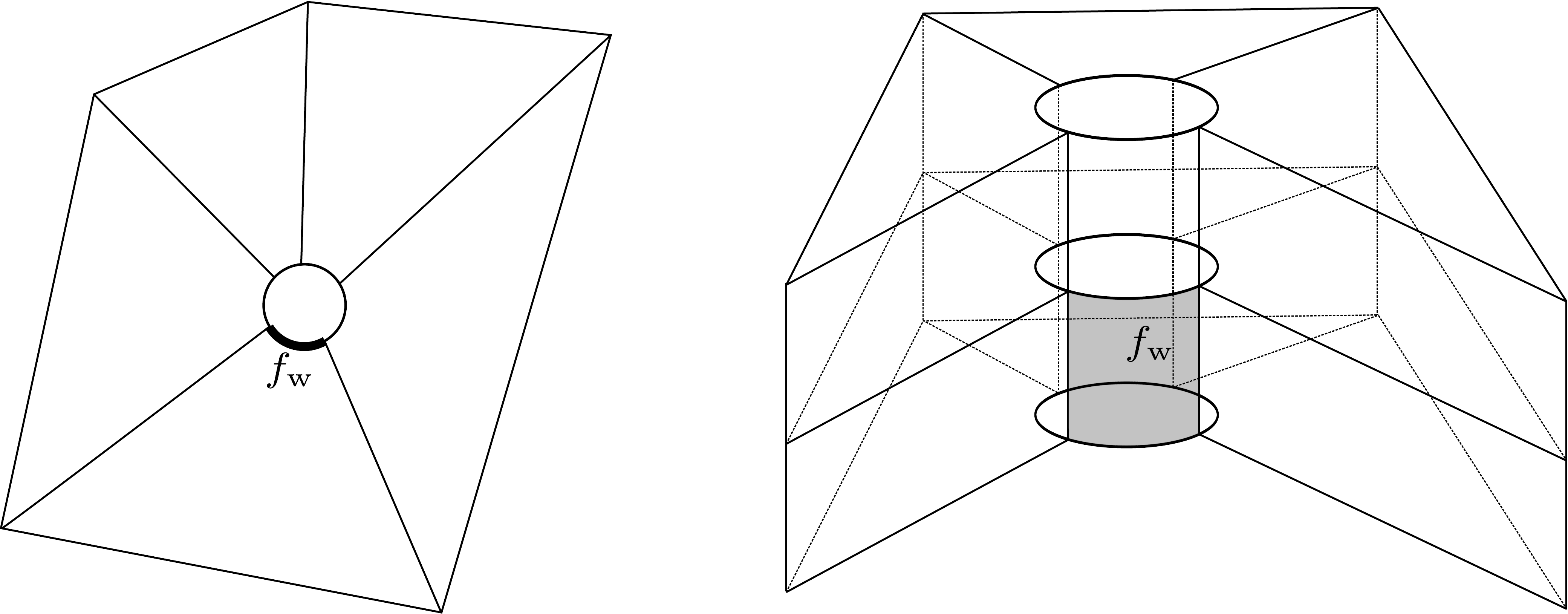}
\caption{Well in two (left) and three (right) dimensions.}
\label{wellDiscret}       
\end{center}
\end{figure}

In commonly available mesh generators it is possible to specify points that are guaranteed to become mesh nodes once the mesh is generated.  Thus we can easily represent a well as a mesh node in two-dimensional models or as an array of mesh edges in three-dimensional models. For a finite volume code, it is more appropriate to associate a well with a cell in two dimensions or with an array of cells in three dimensions. Therefore, we construct cylinders (circles in two dimensions) around well edges (nodes) as in Fig. \ref{wellDiscret}. Another way to represent a well is described in Example \ref{example5}.

%In finite element methods wells are usually associated with arrays of mesh edges in three dimensional models, or with nodes in two dimensional models. Since there are more tools available to generate unstructured meshes for finite elements then for other methods, in order to be able to use these meshes we construct cylinders (circles in two dimensions) around well edges (nodes). As a result, a well is an array of cylindrical cells in three dimensional models (Fig. \ref{wellDiscret} right) or a circular cell in two dimensional models (Fig. \ref{wellDiscret} left). If the well radius is bigger than characteristic length of a mesh cells well is  \hl{Other option also used in this paper, is to model a well as a hole in a model.}

The \textit{well face correction} method (WFC) described in Subsection \ref{Correction} is related to the Peaceman method and results in a greatly improved well extraction rate compared to the uncorrected scheme, but the hydraulic head is still inconsistent even though it is improved. The difference between Peaceman model and WFC is that is that the Peaceman method introduces an additional equation for the well flow rate and does not change the hydraulic head values around the well, while the WFC scheme changes the discretization of flux between two mesh cells which results in an improved hydraulic head accuracy throghout the domain.
In the \textit{near-well correction} scheme (NWC) presented in Section \ref{nearWellScheme}, the flow in a well vicinity is split into a linear part and a part that is due to the influence of the well. This splitting was used in \cite{Din01,Din04} for an otherwise unrelated multipoint scheme, but the accuracy of that scheme reduces if the well is much smaller then the grid size. On the other hand, NWC scheme uses meshes that depend on the well radius, but it is always second-order accurate.

None of these two schemes introduces additional equations or modifies grids apart from introducing well cells, but rather changes the way the flux is approximated on some faces. In the WFC method, the flux through well faces is calculated using a linear two-point approximation. In the NWC method, the flux through faces in the near-well region is approximated using a non-linear two-point approximation. This approximation is obtained as a convex combination of two one-side multipoint linear flux approximations, as in non-linear two-point schemes \cite{Dan09,Lep05,Lip07,Vas08,Vid11,Vid13,Vid14,Yua08}. When compared to these schemes, one-side flux approximations used with the NWC method have different stencils, but the final stencil of the combined flux is the same.

The paper is organized as follows. The two flux discretization schemes are presented in Section \ref{Discr} in the two-dimensional case. Three-dimensional versions of these schemes are presented in Section \ref{3D}. Application of the NWC scheme to a heterogeneous medium is considered in Section \ref{heter}. Results of numerical tests are provided in Section \ref{tests}.

\section{Discretization in two dimensions}
\label{Discr}
In order to use the same terminology in the two-dimensional and three-dimensional cases, the edges of two-dimensional cells are referred to as \textit{faces} and their lengths are called \textit{face areas}. We assume that every cell is a star-shaped set with respect to its barycenter as in \cite{Dan09}.

Integrating (\ref{massCons}) over cell $T$ and applying the divergence theorem yields
\begin{equation}
\sum_{f\in\partial T}\chi_{T,f}u_{f}=\int_{T}g_{\text{s}}{\rm d}T,\quad \text{where}\quad u_{f}=\int_{f}{\bf u}\cdot{\bf n}_{f}{\rm d}s.\label{Discretization}
\end{equation}
Term $u_{f}$ denotes the flux through face $f$, ${\bf n}_{f}$ is a unit vector normal to face $f$ fixed once and for all, while $\chi_{T,f}=1$ if ${\bf n}_{f}$ points outside of $T$ and $\chi_{T,f}=-1$ otherwise. Boundary face normals always point outside. If $\Gamma_{\text{N}}\neq\emptyset$, then we assume that $\bar{\Gamma}_{\text{N}}\cap\bar{\Gamma}_{\text{D}}$ contains only nodes (edges in the three-dimensional case).

We associate one hydraulic head value $h_{T}$ with each cell centroid ${\bf x}_{T}$. The Dirichlet boundary condition is evaluated at each node belonging to $\Gamma_{\text{D}}$. These cell centroids and nodes with associated hydraulic head values are referred to as \textit{primary collocation points}.

An auxiliary hydraulic head value is associated with each well face. These auxiliary head values are eliminated and the hydraulic head is not actually computed there. Since the centroid of a well face $f_{\text{w}}$ does not belong to it because the face is not planar, we define an \textit{auxiliary collocation point} ${\bf x}_{f_{\text{w}}}$ associated with this face as the point on $f_{\text{w}}$ nearest to the centroid.

Either the hydraulic head is set in a well cell or a source/sink term is used in this cell to specify the flow rate. 

\subsection{Well face correction (WFC)}
\label{Correction}
We consider the case of a homogeneous isotropic circular reservoir of radius $R$ with a well of radius $r$ in its center. The well extraction rate \cite{Hai95} is
\begin{equation}
Q=AK\frac{h_{R}-h_{r}}{r\ln \frac{R}{r}},\label{totalFlux}
\end{equation}
where $A$ is the total area of the well screen, while $h_{r}$ and $h_{R}$ are hydraulic head values in the porous medium at distances $r$ and $R$, respectively, from the well center.

Based on the flow rate (\ref{totalFlux}), we propose to calculate the flux through well face $f_{\text{w}}$ (see Fig. \ref{wellDiscret}) belonging to cell $T$ as
\begin{equation}
u_{f_{\text{w}}}=|f_{\text{w}}|K\frac{h_{T}-h_{f_{\text{w}}}}{r\ln \frac{\rho ({\bf x}_{T})}{r}},\label{wellFaceFluxCorr}
\end{equation}
where $|f_{\text{w}}|$ is the face $f_{\text{w}}$ area, ${\bf x}_{T}$ is the centroid of cell $T$, and $\rho ({\bf x}_{T})$ is the distance from ${\bf x}_{T}$ to the well center. Hydraulic head at cell $T$ is denoted by $h_{T}$ and $h_{f_{\text{w}}}$ is the auxiliary hydraulic head value at face $f_{\text{w}}$. 

If the well is not colmated, then $h_{f_{\text{w}}}=h_{\text{w}}$. Otherwise from equation (\ref{colmation}), the flux through face $f_{\text{w}}$ is
\begin{equation}
u_{f_{\text{w}}}=|f_{\text{w}}|\Psi(h_{f_{\text{w}}}-h_{\text{w}}). \label{faceColmation}
\end{equation}
Combining equations (\ref{wellFaceFluxCorr}) and (\ref{faceColmation}) gives a flux approximation that does not include the head value at face $f$:
\begin{equation}
u_{f_{\text{w}}}=|f_{\text{w}}|\frac{\Psi K}{r\Psi\ln\frac{\rho({\bf x}_{T})}{r}+K}(h_{T}-h_{\text{w}}).
\label{wellFaceFlux}
\end{equation}

As shown in Section \ref{tests}, this correction leads to an acceptable well extraction rate.
However, there is a substantial error in the hydraulic head distribution, which does not decrease significantly if the mesh is uniformly refined, unless the mesh is very fine.

\subsection{Near-well correction (NWC)}
\label{nearWellScheme}
The hydraulic head is represented as
\begin{equation}
h\approx L+\hat{h},\label{sum}
\end{equation}
where $L$ is a linear function and $\hat{h}$ is a singular part
\begin{equation}
\hat{h}({\bf x})=C_{0}\ln r({\bf x}),\label{anallyticSol}
\end{equation}
$C_{0}$ is an arbitrary constant, and
\begin{equation}
r({\bf x})=\|{\bf x}-{\bf x}_{\text{w}}\|\label{distFromWell}
\end{equation}
is the distance to the well center ${\bf x}_{\text{w}}$. 

From (\ref{sum}) and (\ref{anallyticSol}), the hydraulic head gradient is
\begin{equation}
\nabla h\approx \nabla L+C_{0}\nabla \left(\ln r({\bf x})\right).
\end{equation}
Thus flux (\ref{Discretization}) can be written as
\begin{equation}
u_{f}=-\int_{f}(K\nabla h)\cdot {\bf n}_{f} {\rm d}s \approx-\int_{f}K\nabla L\cdot {\bf n}_{f}{\rm d}s-
\int_{f}C_{0}K\nabla \left(\ln r({\bf x})\right)\cdot {\bf n}_{f}{\rm d}s.\label{flux}
\end{equation}

Since $\nabla L$ is constant, the first integral can be approximated as
\begin{equation}
-\int_{f}K\nabla L\cdot {\bf n}_{f}{\rm d}s \approx -\left|f\right|K_{f}\nabla L\cdot {\bf n}_{f},\label{fIntegral}
\end{equation}
where $K_{f}=K({\bf x}_{f})$.

\begin{figure}
\begin{center}
  \includegraphics[width=84mm]{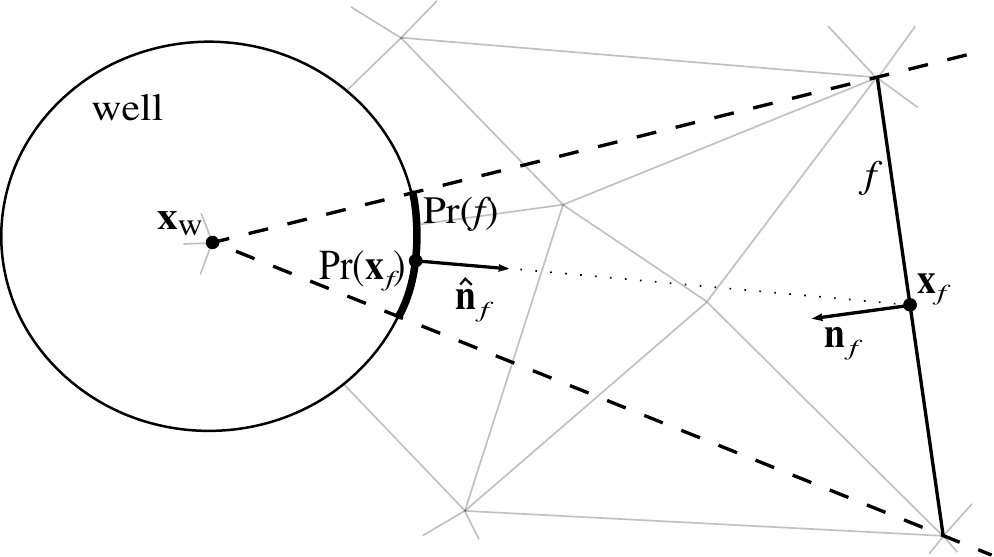}
\caption{Radial projection onto the well.}
\label{faceProjection}       
\end{center}
\end{figure}

Let $\text{Pr}(f)$ denote the radial projection of face $f$ onto the well wall from the well center (see Fig. \ref{faceProjection}). The flow component described by the second integral is directed toward the well center. Therefore, this flux component through face $f$ is the same as through $\text{Pr}(f)$:  
\begin{equation}
-\int_{f}C_{0}K\nabla(\ln r({\bf x}))\cdot{\bf n}_{f}{\rm d}s\approx -K_{f}\int_{\text{Pr}(f)}
C_{0}\nabla(\ln r(\text{Pr}({\bf x})))\cdot \hat{{\bf n}}_{f}{\rm d}s= -\sigma_{f}\frac{\left| \text{Pr}(f)\right|C_{0}K_{f}}{r},\label{sIntegral}
\end{equation}
because
\begin{equation}
\nabla\left(\ln r(\text{Pr}({\bf x}))\right)\cdot\hat{\bf n}_{f}=\frac{1}{r},
\end{equation}
where $\hat{{\bf n}}_{f}$ is the outer unit normal to the circle at Pr$({\bf x})$, and $\sigma_{f}=-1$ if ${\bf n}_{f}$ points inside the triangle defined by face $f$ and the well center, or $\sigma_{f}=1$ otherwise. 
Substituting (\ref{fIntegral}) and (\ref{sIntegral})  in (\ref{flux}) gives
\begin{equation}
u_{f}\approx u_{f,T}=-\left|f\right|(K_{f}\nabla L_{f,T})\cdot {\bf n}_{f}-\sigma_{f}\frac{\left| \text{Pr}(f)\right|C_{0}K_{f}}{r},\label{fluxA}
\end{equation}
where cell $T$ contains face $f$. We have added this subscript because we associate one approximation (\ref{fluxA}) with each cell-face pair.

We use (\ref{faceColmation}) and (\ref{fluxA}) to express the auxiliary hydraulic head values $h_{f_{\text{w}}}$. Let ${\bf n}_{f_{\text{w}}}$ be directed outside of the well. Then, from (\ref{faceColmation}) and (\ref{fluxA})
\begin{equation}
|f_{\text{w}}|\Psi (h_{f_{\text{w}}}-h_{\text{w}})=-|f_{\text{w}}|(K_{f_{\text{w}}}\nabla L)\cdot{\bf n}_{f_{\text{w}}}-\sigma_{f_{\text{w}}}\frac{\left|\text{Pr}(f_{\text{w}})\right|C_{0}K_{f_{\text{w}}}}{r}.
\end{equation}
From this equation we express $h_{f_{\text{w}}}$:
\begin{equation}
h_{f_{\text{w}}}=h_{\text{w}}-\frac{(K_{f_{\text{w}}}\nabla L)\cdot {\bf n}_{f_{\text{w}}}}{\Psi}-
\sigma_{f_{\text{w}}}\frac{\left|\text{Pr}(f_{\text{w}})\right|C_{0}K_{f_{\text{w}}}}{\Psi r |f_{\text{w}}|}.\label{headWellFace}
\end{equation}

Let ${\bf x}_{i}$ be a collocation point other than ${\bf x}_{T}$. The difference between the hydraulic head values at ${\bf x}_{i}$ and ${\bf x}_{T}$ is
\begin{equation}
h_{i}-h_{T}\approx \nabla L\cdot({\bf x}_{i}-{\bf x}_{T})+C_{0}\ln\frac{r({\bf x}_{i})}{r({\bf x}_{T})}.\label{diff}
\end{equation}
We would like to determine $\nabla L$ and $C_{0}$ so that a set of conditions such as (\ref{diff}) is satisfied.

If ${\bf x}_{i}$ is a well face auxiliary collocation point, then (\ref{diff}) becomes
\begin{equation}
h_{f_{\text{w}}}-h_{T}\approx \nabla L\cdot({\bf x}_{f_{\text{w}}}-{\bf x}_{T})+C_{0}\ln\frac{r({\bf x}_{f_{\text{w}}})}{r({\bf x}_{T})}.
\label{diffBun}
\end{equation}
By substituting (\ref{headWellFace}) in this relation we eliminate $h_{f_{\text{w}}}$:
\begin{equation}
h_{\text{w}}-h_{T}\approx \nabla L\cdot\left({\bf x}_{f_{\text{w}}}-{\bf x}_{T}+\frac{K_{f_{\text{w}}}}{\Psi}{\bf n}_{f}\right)+C_{0}\left(\ln \frac{r({\bf x}_{f_{\text{w}}})}{r({\bf x}_{T})}+\sigma_{f_{\text{w}}}\frac{\left|\text{Pr}(f_{\text{w}})\right|C_{0}K_{f_{\text{w}}}}{\Psi r |f_{\text{w}}|}\right).
\label{kaBunaru}
\end{equation}

On a Neumann boundary face $\bar{f}$ we may require that the flux computed using formula (\ref{flux}) satisfies the boundary condition. The corresponding equation is obtained by integrating (\ref{Neumann}) over $\bar{f}$ and using (\ref{fluxA}): 
\begin{equation}
\left|\bar{f}\right|(K_{\bar{f}}\nabla L)\cdot {\bf n}_{\bar{f}}+\sigma_{f}\frac{\left| \text{Pr}(\bar{f})\right|C_{0}K_{\bar{f}}}{r}
=-g_{\text{N}}({\bf x}_{\bar{f}})|\bar{f}|.\label{NeumB}
\end{equation}

If $T$ is a well cell, then we use a well face auxiliary collocation point instead of ${\bf x}_{T}$ in (\ref{diff}):
\begin{equation}
h_{i}-h_{f_{\text{w}}}\approx \nabla L\cdot({\bf x}_{i}-{\bf x}_{f_{\text{w}}})+C_{0}\ln\frac{r({\bf x}_{i})}{r({\bf x}_{f_{\text{w}}})}.\label{diffWell}
\end{equation}
Substituting (\ref{headWellFace}) in this relation we eliminate: $h_{f_{\text{w}}}$
\begin{equation}
h_{i}-h_{\text{w}}\approx \nabla L\cdot\left({\bf x}_{i}-{\bf x}_{f_{\text{w}}}-\frac{K_{f_{\text{w}}}}{\Psi}{\bf n}_{f}\right)+C_{0}\left(\ln \frac{r({\bf x}_{i})}{r({\bf x}_{f_{\text{w}}})}-\sigma_{f_{\text{w}}}\frac{\left|\text{Pr}(f_{\text{w}})\right|C_{0}K_{f_{\text{w}}}}{\Psi r |f_{\text{w}}|}\right).
\label{odBunara}
\end{equation}

To approximate the flux using (\ref{fluxA}) we need to determine 
\begin{equation}
{\bf C}=\left[\frac{\partial L}{\partial x}\quad \frac{\partial L}{\partial y} \quad C_{0}\right]^{\text{T}}.
\label{nepoz}
\end{equation}
These are found by solving a linear system
\begin{equation}
A{\bf C}\approx{\bf b}\label{matr}
\end{equation}
consisting of three equations of type (\ref{diff}), (\ref{kaBunaru}), (\ref{NeumB}), or appropriately transformed equations such as (\ref{odBunara}) when $T$ is a well cell.

If matrix $A$ is not invertible then other equations of form (\ref{diff}), (\ref{kaBunaru}), (\ref{NeumB}), or appropriately transformed equations such as (\ref{odBunara}) are chosen to form system (\ref{matr}). 

Let us assume that matrix $A$ is invertible, and let elements of matrix $A^{-1}$ be denoted by $a_{ij}$. Let index $k$ correspond to collocation points in equations of form (\ref{diff}) and (\ref{kaBunaru}) (or (\ref{odBunara})), while index $\bar{k}$ corresponds to Neumann boundary faces in equations of form (\ref{NeumB}).
From (\ref{matr}), the coordinates of the unknown vector ${\bf C}$ are:
\[
\frac{\partial L}{\partial x}\approx\sum_{k}a_{1k}(h_{k}-h_{T})-\sum_{\bar{k}}a_{1\bar{k}}g_{\text{N}}({\bf x}_{f_{\bar{k}}})|f_{\bar{k}}|,
\]
\begin{equation}
\frac{\partial L}{\partial y}\approx\sum_{k}a_{2k}(h_{k}-h_{T})-\sum_{\bar{k}}a_{2\bar{k}}g_{\text{N}}({\bf x}_{f_{\bar{k}}})|f_{\bar{k}}|,\label{vecX}
\end{equation}
\[
C_{0}\approx\sum_{k}a_{3k}(h_{k}-h_{T})-\sum_{\bar{k}}a_{3\bar{k}}g_{\text{N}}({\bf x}_{f_{\bar{k}}})|f_{\bar{k}}|.
\]

After substituting (\ref{vecX}) in (\ref{fluxA}), the flux approximation becomes 
\begin{equation}
u_{f}\approx-\sum_{k}\alpha_{k}(h_{k}-h_{T})+ \sum_{\bar{k}} \alpha_{\bar{k}}g_{\text{N}}({\bf x}_{f_{\bar{k}}})|f_{\bar{k}}|,\label{fluxApp}
\end{equation}
where 
\begin{equation}
\alpha_{k}=
K_{f}\left(|f|\left(a_{1k}n_{f}^{1} + a_{2k}n_{f}^{2}\right) + a_{3k}\sigma_{f}\frac{|\text{Pr}(f)|}{r}\right),\label{alpha} 
\end{equation}
\begin{equation}
\alpha_{\bar{k}}=K_{f}\left(|f|\left(a_{1\bar{k}}n_{f}^{1} + a_{2\bar{k}}n_{f}^{2}\right)+ a_{3\bar{k}}\sigma_{f}\frac{|\text{Pr}(f)|}{r}\right).\label{alphaNad}
\end{equation}
Term $n_{f}^{l}$ denotes the $l$-th coordinate of vector ${\bf n}_{f}$. 

Let cells $T_{+}$ and $T_{-}$ share face $f$, and let ${\bf n}_{f}$ point from $T_{+}$ to $T_{-}$. One-side approximations (\ref{fluxApp}) of the flux through face $f$ from cell $T_{+}$ or $T_{-}$ are, respectively,
\begin{equation}
u_{f}\approx u_{f,+}=-\sum_{k}\alpha_{k}(h_{k}-h_{+})+ \sum_{\bar{k}} \alpha_{\bar{k}}g_{\text{N}}({\bf x}_{f_{\bar{k}}})|f_{\bar{k}}|,
\end{equation}
\begin{equation}
u_{f}\approx -u_{f,-} = \sum_{l}\alpha_{l}(h_{l}-h_{-})- \sum_{\bar{l}} \alpha_{\bar{l}}g_{\text{N}}({\bf x}_{f_{\bar{l}}})|f_{\bar{l}}|,\label{fluxApp2}
\end{equation}

The derivation is carried out further as in \cite{Dan09,Vid11,Vid14}.
One-side approximations of form (\ref{fluxApp}) and (\ref{fluxApp2}) are linearly combined using non-negative weights $\mu_{+}$ and $\mu_{-}$:
\begin{equation}
u_{f} \approx -\mu_{+}\sum_{k}\alpha_{k}^{+}(h_{k}-h_{+}) + \mu_{-} \sum_{l}\alpha_{l}^{-}(h_{l}-h_{-}) +
\mu_{+}\sum_{\bar{k}}\alpha_{\bar{k}}^{+}g_{\text{N}}({\bf x}_{f_{\bar{k}}})|f_{\bar{k}}|-\mu_{-}\sum_{\bar{l}}\alpha_{\bar{l}}^{-}g_{\text{N}}({\bf x}_{f_{\bar{l}}})|f_{\bar{l}}|.
\label{faceFlux}
\end{equation}
For this approximation to be valid, it is required that
\begin{equation}
\mu_{+}+\mu_{-}=1.\label{mi}
\end{equation}
We choose $\mu_{+}$ and $\mu_{-}$ so that in (\ref{faceFlux}) the contributions of hydraulic head values other then $h_{-}$ and $h_{+}$, as well as the contributions of inflow Neumann boundary conditions, cancel out: 
\begin{equation}
-\mu_{+}d_{+}+\mu_{-}d_{-}=0,\quad d_{\pm}=\sum_{\substack{k \\ \mathbf x_k\neq\mathbf x_\mp}}
\alpha_{k}^{\pm}h_{k} - 
\sum_{\substack{\bar{k} \\g_{\text{N}}({\bf x}_{f_{\bar{k}}})<0}}
\alpha_{\bar{k}}^{\pm}g_{\text{N}}({\bf x}_{f_{\bar{k}}})|f_{\bar{k}}|.\label{mid}
\end{equation}

If $d_{+}+d_{-}\neq 0$, $\mu_{\pm}$ is computed from (\ref{mi}) and (\ref{mid}) as
\begin{equation}
\mu_{+}=\frac{d_{-}}{d_{+}+d_{-}}, \quad \mu_{-}=\frac{d_{+}}{d_{+}+d_{-}},
\end{equation}
otherwise we set $\mu_{\pm}=0.5$.

In this way, a two-point flux approximation is obtained:
\begin{equation}
u_{f}\approx M_{f}^{+}h_{+}-M_{f}^{-}h_{-}+r_{f}\label{twopoint},
\end{equation}
where
\begin{equation}
M_{f}^{+}=\mu_{+}\sum_{k}\alpha_{k}^{+}+\mu_{-}\sum_{\substack{l\\{\bf x}_{l}={\bf x}_{+}}}\alpha_{l}^{-}, 
\end{equation}
\begin{equation}
M_{f}^{-}=\mu_{-}\sum_{l}\alpha_{l}^{-}+\mu_{+}\sum_{\substack{k\\{\bf x}_{k}={\bf x}_{-}}}\alpha_{k}^{+}, 
\end{equation}
\begin{equation}
r_{f}=\mu_{+}\sum_{\substack{\bar{k} \\g_{\text{N}}({\bf x}_{f_{\bar{k}})}>0}} \alpha_{\bar{k}}^{+}g_{\text{N}}({\bf x}_{f_{\bar{k}}})|f_{\bar{k}}|-\mu_{-}\sum_{\substack{\bar{l} \\g_{\text{N}}({\bf x}_{f_{\bar{l}})}>0}} \alpha_{\bar{l}}^{-}g_{\text{N}}({\bf x}_{f_{\bar{l}}})|f_{\bar{l}}|.
\end{equation}

Using this approximation in (\ref{Discretization}) in the near-well region and scheme \cite{Vid14} outside of this region, we obtain a system of equations 
\begin{equation}
\mathcal{A}({\bf h}){\bf h}=\mathfrak{b}({\bf h}).
\end{equation}
This system is non-linear because $M_{f}^{\pm}$ depends on the discrete hydraulic head values $h_{i}$ through $\mu_{\pm}$ and $d_{\pm}$. It can be linearized using Picard method:
\begin{equation}
\mathcal{A}({\bf h}^{n}){\bf h}^{n+1}=\mathfrak{b}({\bf h}^{n}).
\end{equation}
Starting with some initial solution ${\bf h}^{0}$, each succeeding iteration is found using a linear solver until the convergence criterion
\begin{equation}
r({\bf h}^{n})=\frac{\|\mathcal{A}({\bf h}^{n}){\bf h}^{n}-\mathfrak{b}({\bf h}^{n})\|}{\|\mathfrak{b}({\bf h}^{n})\|}<\varepsilon
\label{PikarStop}
\end{equation}
is met for a small $\varepsilon$ set in advance, or until the maximal number of iterations is reached.

Following the same logic as in \cite{Dan09}, it is required that $\alpha_{k}, \alpha_{\bar{k}}\geq 0$ for all $k,\bar{k}$, which implies that $M_{f}^{\pm}\geq 0$, so the resulting computational matrix $\mathcal{A}({\bf h})$ is an M-matrix and the method preserves the solution positivity. If this is not the case, then other equations of form (\ref{diff}), (\ref{kaBunaru}), (\ref{NeumB}), or transformed equations such as (\ref{odBunara}), are chosen to form (\ref{matr}). 

\begin{figure}[!htbp]
\centering
   \includegraphics[width=152mm]{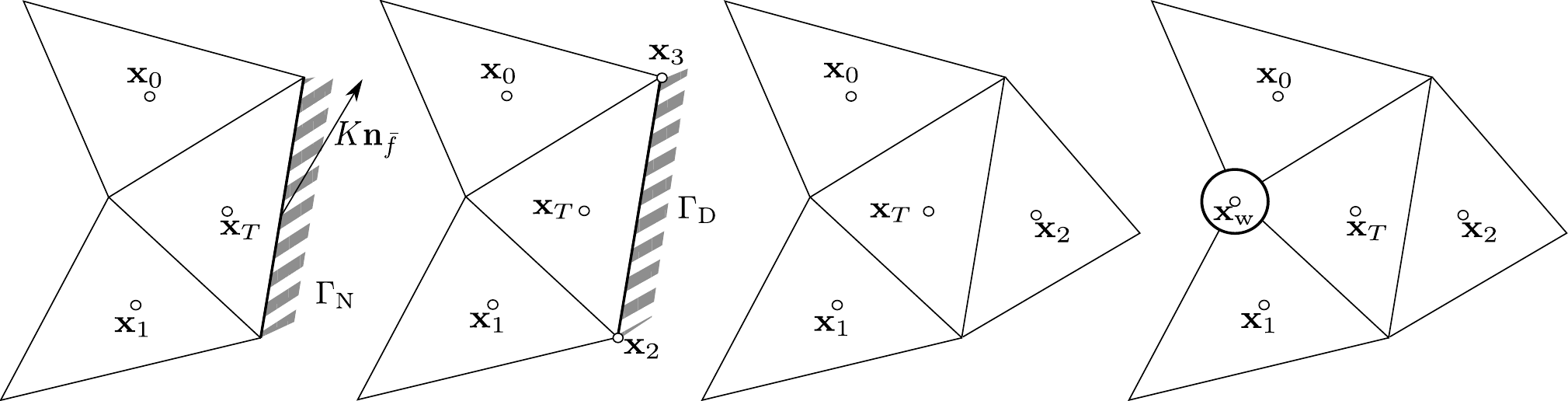}\\
\caption{Initial set of collocation points and Neumann boundary faces for cell $T$ when: one of its faces belongs to Neumann boundary (first), one of its faces belongs to Dirichlet boundary (second), there are no boundary faces (third), and faces of $T$ is a well face (fourth).}
\label{colloc}       
\end{figure}

\begin{figure}[!htbp]
\centering
   \includegraphics[width=100mm]{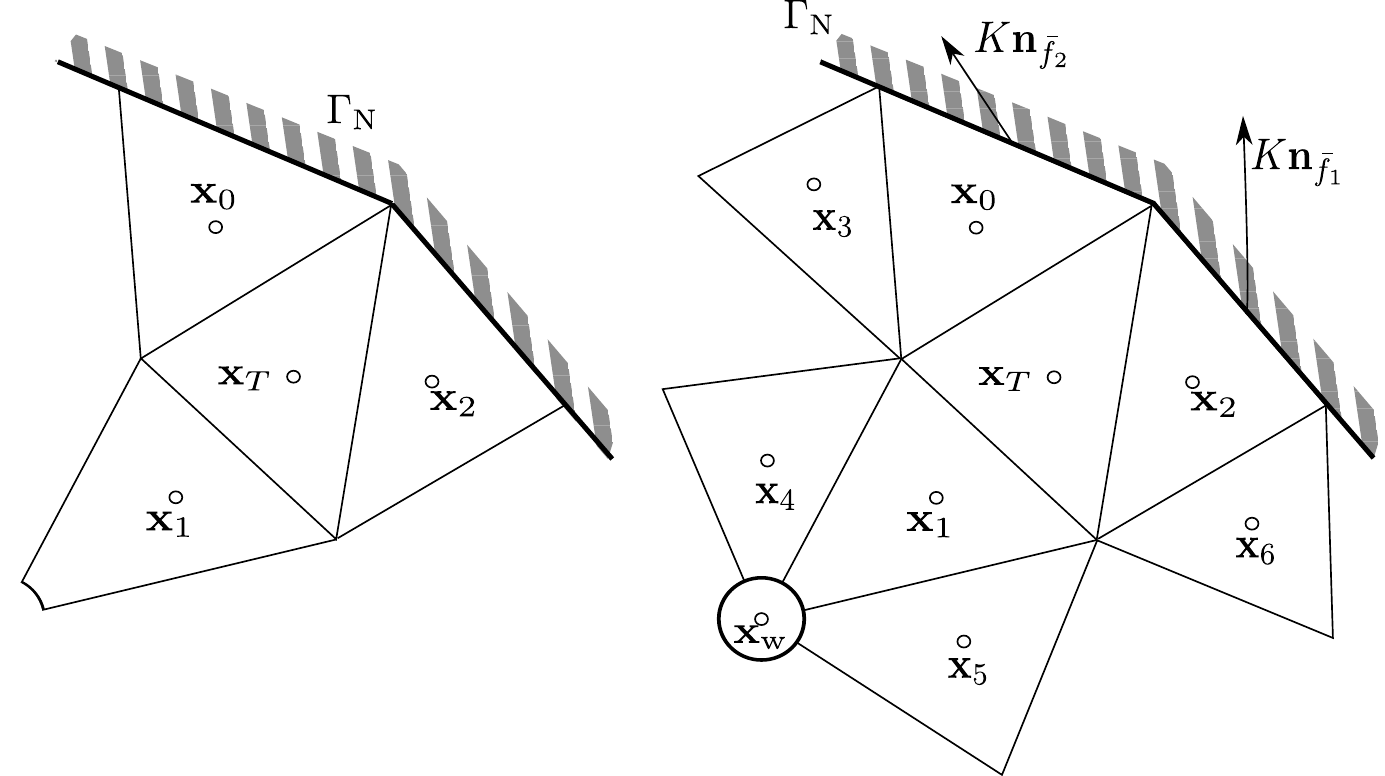}\\
\caption{Initial set of collocation points (left) and its first expansion (right).}
\label{expand}       
\end{figure}

The search for these equations is performed by testing all combinations of cells and boundary conditions belonging to a {\it candidate set}. Initially, the candidate set consist of cells that share a face with $T$, Neumann boundary faces of $T$, Dirichlet boundary nodes of $T$, and well faces of $T$. Examples of initial candidate sets are shown in Fig. \ref{colloc}. If every combination of equations corresponding to elements of the candidate set leads to negative $\alpha_{k}$ or $\alpha_{\bar{k}}$, then the candidate set is expanded by adding all neighbouring cells, Neumann boundary faces, Dirichlet boundary nodes, and well faces of all cells already in the candidate set. After the candidate set expansion, we again test all combinations of its elements. This process is repeated until a set of non-negative $\alpha_{k}$ and $\alpha_{\bar{k}}$ is obtained. Example of initial set expansion is shown in Fig. \ref{expand}.

In practice, the candidate set is rarely expanded more than once. Although we cannot prove that a set of non-negative $\alpha_{k}$ and $\alpha_{\bar{k}}$ can always be found, in our practice we have not encountered a case where this would not be so. Nevertheless, there is an artificially constructed example in \cite{Vid13} where non-negative coefficients could not be found in a simpler case that does not include wells. 

The proposed scheme is used within a near-well region, which can be of any shape, as long as it includes at least the cells nearest to the well. Near-well regions belonging to different wells must not overlap. Scheme \cite{Vid14} is used outside of these regions. Fluxes through faces between the near-well region and the rest of the model are obtained by coupling the one-side flux approximation (\ref{fluxApp}) with the uncorrected one-side flux approximation used in \cite{Vid14} in the same way as in (\ref{faceFlux}).

As in the two-point non-linear scheme \cite{Dan09,Vid11,Vid14}, the convergence of the NWC method cannot be shown analytically, but numerical examples (see Section \ref{tests}) indicate that it is second-order accurate.

\section{Discretization in three dimensions}
\label{3D}
A well is represented as an array of cylindrical cells in three dimensions. Either the hydraulic head or a source term is specified in the well cell where the pump is located. The flow through the well is modeled using the Hagen-Poiseuille law \cite{Sut93}, meaning that the hydraulic conductivity along the well is computed as
\begin{equation}
K_{\text{w}}=\frac{r^{2}\rho g}{8\mu},
\end{equation}
where $\rho$ is the density, $g$ is the standard gravity, and $\mu$ is the dynamic viscosity.

The WFC scheme derived in Section \ref{Correction} is directly applicable to the three-dimensional case.

The NWC scheme is derived in a manner analogous to the two-dimensional case, with $\rho ({\bf x}_{T})$ representing the distance between ${\bf x}_{T}$ and the well axis, and Pr$(f)$ denoting a projection of face $f$ onto the well cylinder. This projection, defined in \ref{AppendixB}, is known in cartography as Lambert cylindrical equal-area projection. 

System (\ref{matr}) is formed with four instead of three equations of type (\ref{diff}), (\ref{kaBunaru}), (\ref{NeumB}), or (\ref{odBunara}). The vector of unknowns is
\begin{equation}
{\bf C}=\left[\frac{\partial L}{\partial x}\quad \frac{\partial L}{\partial y} \quad\frac{\partial L}{\partial z}\quad C_{0}\right]^{T}.
\end{equation}
Thus, instead of (\ref{vecX}) we have
\[
\frac{\partial L}{\partial x}\approx\sum_{k}a_{1k}(h_{k}-h_{0})-\sum_{\bar{k}}a_{1\bar{k}}g_{\text{N}}({\bf x}_{f_{\bar{k}}})|f_{\bar{k}}|,
\]
\begin{equation}
\frac{\partial L}{\partial y}\approx\sum_{k}a_{2k}(h_{k}-h_{0})-\sum_{\bar{k}}a_{2\bar{k}}g_{\text{N}}({\bf x}_{f_{\bar{k}}})|f_{\bar{k}}|,\label{vecX1}
\end{equation}
\[
\frac{\partial L}{\partial z}\approx\sum_{k}a_{3k}(h_{k}-h_{0})-\sum_{\bar{k}}a_{3\bar{k}}g_{\text{N}}({\bf x}_{f_{\bar{k}}})|f_{\bar{k}}|,
\]
\[
C_{0}\approx\sum_{k}a_{4k}(h_{k}-h_{0})-\sum_{\bar{k}}a_{4\bar{k}}g_{\text{N}}({\bf x}_{f_{\bar{k}}})|f_{\bar{k}}|.
\]
Therefore, instead of (\ref{alpha}) and (\ref{alphaNad}) we have
\begin{equation}
\alpha_{k}=
K_{f}\left(|f|\sum_{l=1}^{3}a_{lk}n_{f}^{l} + a_{3k}\sigma_{f}\frac{|\text{Pr}(f)|}{r}\right),\label{alpha1} 
\end{equation}
and 
\begin{equation}
\alpha_{\bar{k}}=K_{f}\left(|f|\sum_{l=1}^{3}a_{l\bar{k}}n_{f}^{l} + a_{3\bar{k}}\sigma_{f}\frac{|\text{Pr}(f)|}{r}\right).\label{alphaNad1}
\end{equation}

\section{Heterogeneous case}
\label{heter}
We tested these schemes in the case of continuous heterogeneous porous media as well as in the discontinuous case. The results presented in Examples \ref{example5} and \ref{example4} show that in the continuous case, as well as in the homogeneous case, the NWC scheme is second-order accurate. 

If the porous medium is discontinuous, then we assume that discontinuities occur only at mesh faces. We can distinguish two cases. In the first case the discontinuity passes away from the well. In this case, the near-well zone for the NWC scheme should include only cells in a single material zone.

In the second case the discontinuity passes through the well center. 
It is impossible to construct an accurate flux discretization with a discontinuity using only hydraulic head at collocation points in a single material zone. Therefore, we apply the piecewise linear transformation introduced in \cite{Vid13, Vid14}. Thus, instead of (\ref{diff}) we have
\begin{equation}
h_{i}-h_{+}\approx \nabla L\cdot F({\bf x}_{i})+C_{0}\ln\frac{r({\bf x}_{i})}{r({\bf x}_{+})},
\end{equation}
and instead of (\ref{NeumB}) 
\begin{equation}
\left|\bar{f}\right|{\bf n}_{\bar{f}}^{T}\nabla F({\bf x}_{\bar{f}})(K_{f}({\bf x}_{\bar{f}})\nabla L)
+\sigma_{f}\frac{\left| \text{Pr}(\bar{f})\right|C_{0}K_{f}}{r}
=-g_{N}({\bf x}_{\bar{f}})|\bar{f}|,
\end{equation}
where $F$ is the piecewise linear transformation depending on the hydraulic conductivity and geometry but not on the hydraulic head. For details of this transformation see \cite{Vid13,Vid14}. Otherwise, the scheme is constructed as in Subsection \ref{nearWellScheme}. The results obtained in Example \ref{example6} show that the NWC scheme remains second-order accurate.

\section{Numerical tests}
\label{tests}
To verify the schemes, we solve several problems (Examples \ref{example1}, \ref{example2}, \ref{example21}, \ref{example3}, \ref{example6}) whose analytical solutions are available. In each of these examples we compare the analytical solution to the results obtained with the uncorrected, WFC, and NWC schemes. In the heterogeneous case, the analytical solution is not available (Examples \ref{example5} and \ref{example4}), so instead of the exact solution we use the solution obtained on the finest mesh. We use natural neighbor interpolation \cite{Sib81} to interpolate this solution to coarser meshes.
Near-well regions are taken to be circular or cylindrical in all examples.

The meshes used in the examples were independently generated and are not hierarchically related. Mesh parameter $P$ is the square root of the largest cell area in the two-dimensional cases (Examples \ref{example1}, \ref{example2}, \ref{example5}, \ref{example4} and \ref{example6}). In the three-dimensional case (Example \ref{example3}), the mesh parameter is the cubic root of the largest cell volume. Unstructured triangular meshes are used in all examples except in third example where unstructured triangular prismatic meshes are used.

The weighted discrete $L_{2}$ and maximum norms are used to evaluate relative hydraulic head errors:
\begin{equation}
\epsilon_{2}^{h}=\left[ \frac{\sum_{T}(h({\bf x}_{T})-h_{T})^{2}|T|}{\sum_{T}(h({\bf x}_{T}))^{2}|T|}\right]^{1/2},
\label{l2Error}
\end{equation}
\begin{equation}
\epsilon_{\max}^{h}=\frac{\max_{T}|h({\bf x}_{T})-h_{T}|}{\left[ \sum_{T}(h({\bf x}_{T}))^{2}|T|/\sum_{T}|T|\right]^{1/2}},
\label{maxError}
\end{equation}
where $|T|$ stands for the volume (area in 2D) of cell $T$. The exact hydraulic head evaluated at the centroid of cell $T$ is denoted by $h({\bf x}_{T})$, while the head value numerically obtained in this cell is denoted by $h_{T}$. These two quantities were scaled with the same value in order to that the weighted discrete $L_{2}$ norm is less or equal to the weighted maximum norm with the equality holding for constant vectors.

The relative error of the total well flux is computed as:
\begin{equation}
\epsilon_{Q}=\frac{Q-Q_{A}}{Q_{A}},
\end{equation}
where $Q$ is the numerical well flux and $Q_{A}$ is the analytical flux. 

The number of Picard iterations needed to obtain the results for $\epsilon=10^{-12}$ in (\ref{PikarStop}) is denoted by $n_{\text{Pic}}$. We take ${\bf h}^{0}={\bf 0}$ for the initial solution in all tests.

\begin{example}
\label{example1}
We consider a circular reservoir with a well in the center $(0,0)$. The exact flow rate is given by (\ref{totalFlux}), and the exact hydraulic head at distance $\rho$ from the center is
\begin{equation}
h(\rho)=\frac{h_{r}\ln\frac{R}{\rho}+h_{R}\ln\frac{\rho}{r}}{\ln\frac{R}{r}}.
\label{pr1}
\end{equation}
In this example we specify the hydraulic head in the well $h_{\text{w}}$ and at $\rho=R$. We take $R=200$, $h_{\text{w}}=55$, $h_{R}=100$ and $K=0.0001$. Transfer coefficient $\Psi$ is set so that the hydraulic head at the well wall is $h_{r}=60$.

\begin{table}[!htbp]
\caption{Errors for the well radius $r=0.05$ in Example \ref{example1}.}
\label{table1}   
\begin{center}
\begin{tabular}{c c c c c c c}
\hline
$P$ & 32$\sqrt{2}$ & 16$\sqrt{2}$ & 8$\sqrt{2}$ & 4$\sqrt{2}$ & 2$\sqrt{2}$ & $\sqrt{2}$ \\
\hline
\multicolumn{7}{l}{Uncorrected scheme}\\
$\epsilon_{2}^{h}$ & 8.46e-02 & 6.65e-02 & 4.61e-02 & 3.61e-02 & 2.67e-02 & 1.93e-02 \\
$\epsilon_{\max}^{h}$  & 2.25e-01 & 2.20e-01 & 2.05e-01 & 1.87e-01 & 1.65e-01 & 1.37e-01\\
$\epsilon_{Q}$ & 2.33e+00 & 1.78e+00 & 1.22e+00 & 9.59e-01 & 7.12e-01 & 5.17e-01 \\
$n_{\text{Pic}}$ & 8 & 9 & 11 & 11 & 12 & 11\\
\\
\multicolumn{7}{l}{WFC scheme}\\
$\epsilon_{2}^{h}$ & 1.37e-03 & 8.46e-04 & 6.36e-04 & 4.54e-04 & 4.29e-04 & 4.86e-04\\
$\epsilon_{\max}^{h}$ & 6.48e-03 & 7.64e-03 & 7.12e-03 & 6.41e-03 & 6.33e-03 & 6.65e-03\\
$\epsilon_{Q}$  & 9.60e-03 & 1.17e-02 & 1.01e-02 & 1.18e-02 & 1.12e-02 & 1.31e-02\\
$n_{\text{Pic}}$ & 8 & 9 & 11 & 11 & 12 & 11\\
\\
\multicolumn{7}{l}{NWC scheme}\\
$\epsilon_{2}^{h}$ & 7.65e-04 & 2.73e-04 & 5.62e-05 & 1.03e-05 & 2.40e-06 & 6.27e-07\\
$\epsilon_{\max}^{h}$ & 3.37e-03 & 1.94e-03 & 6.05e-04 & 9.33e-05 & 2.67e-06 & 8.39e-07\\
$\epsilon_{Q}$ & 4.42e-03 & -1.56e-03 & 1.73e-04 & 4.93e-05 & 6.52e-06 & 2.06e-06\\
$n_{\text{Pic}}$ & 8 & 9 & 11 & 11 & 12 & 12\\
\noalign{\smallskip}\hline
\end{tabular}
\end{center}
\end{table}

The errors are presented in Table \ref{table1}. The uncorrected scheme is inconsistent in the maximum norm for the considered meshes and the flow rate through the well is completely wrong. The hydraulic head error is larger near the well, as shown in Fig. \ref{headError1} (left). This is as expected because the flow velocity changes quickly in this region.

If the WFC scheme is used, the errors are smaller than those obtained without any correction. The largest errors are still located near the well (Fig. \ref{headError1}, middle). The well flow rate error is around one percent on the coarsest mesh and it does not decrease as the mesh is refined. Therefore, the scheme is inconsistent.

The results for the NWC scheme were obtained using a near-well region with radius $40$. The absolute hydraulic head error distribution is shown in Fig. \ref{headError1} (right).
The results obtained in this way are second-order accurate. If we took $R$ for the radius of the near-well region, then this scheme would be exact.

The reduction of the well flow rate error with the mesh parameter is less predictable because it depends on the particular geometry of the few cells around the well, which changes in a random fashion as the mesh is refined. Nevertheless, a comparison of flow rate errors on the finest and coarsest meshes shows that this flow rate is at least first-order accurate.

\begin{figure}[!htbp]
\centering
   \begin{tabular}{@{}ccc@{}}
   \includegraphics[width=52mm]{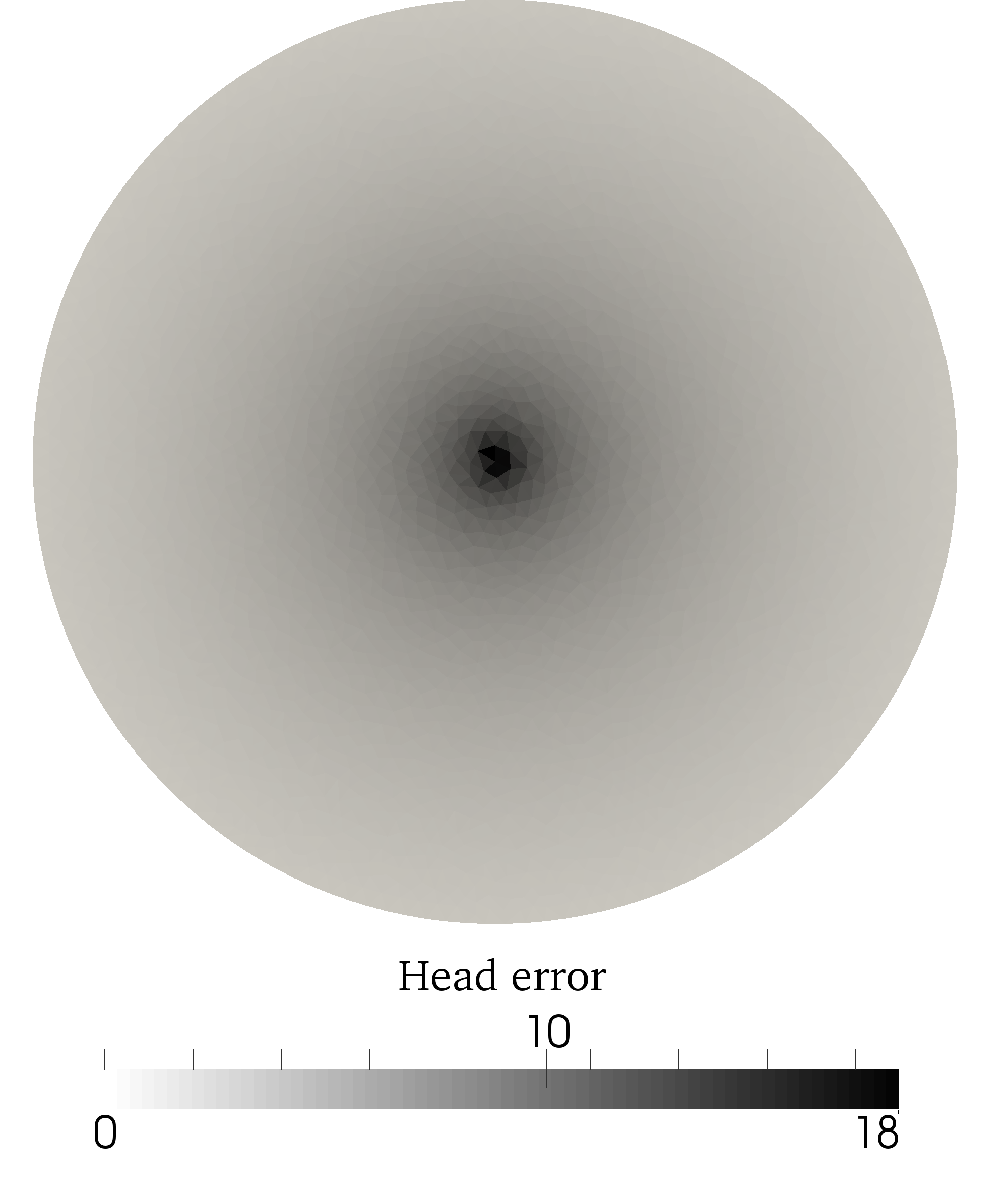}&
   \includegraphics[width=52mm]{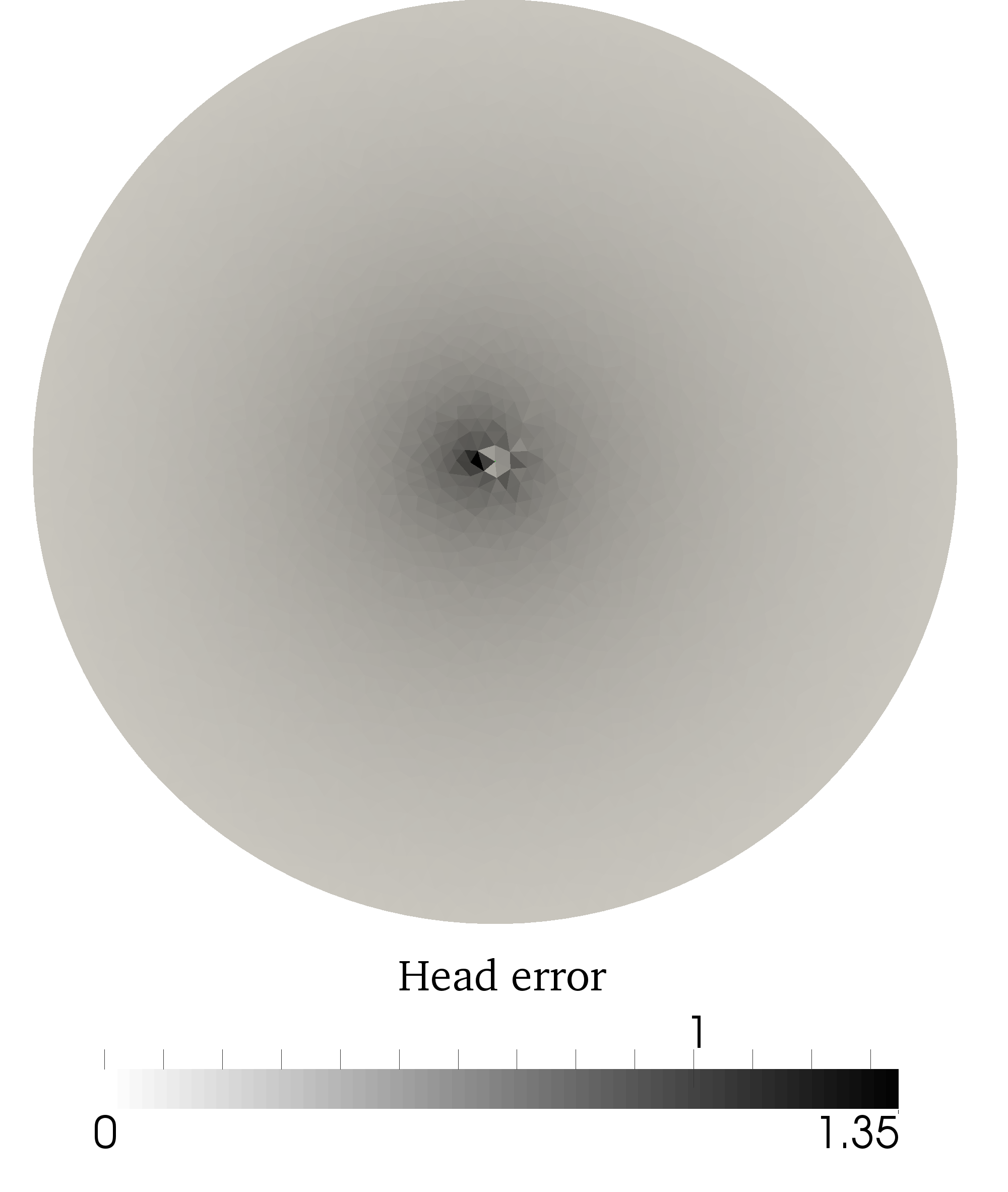}&
   \includegraphics[width=52mm]{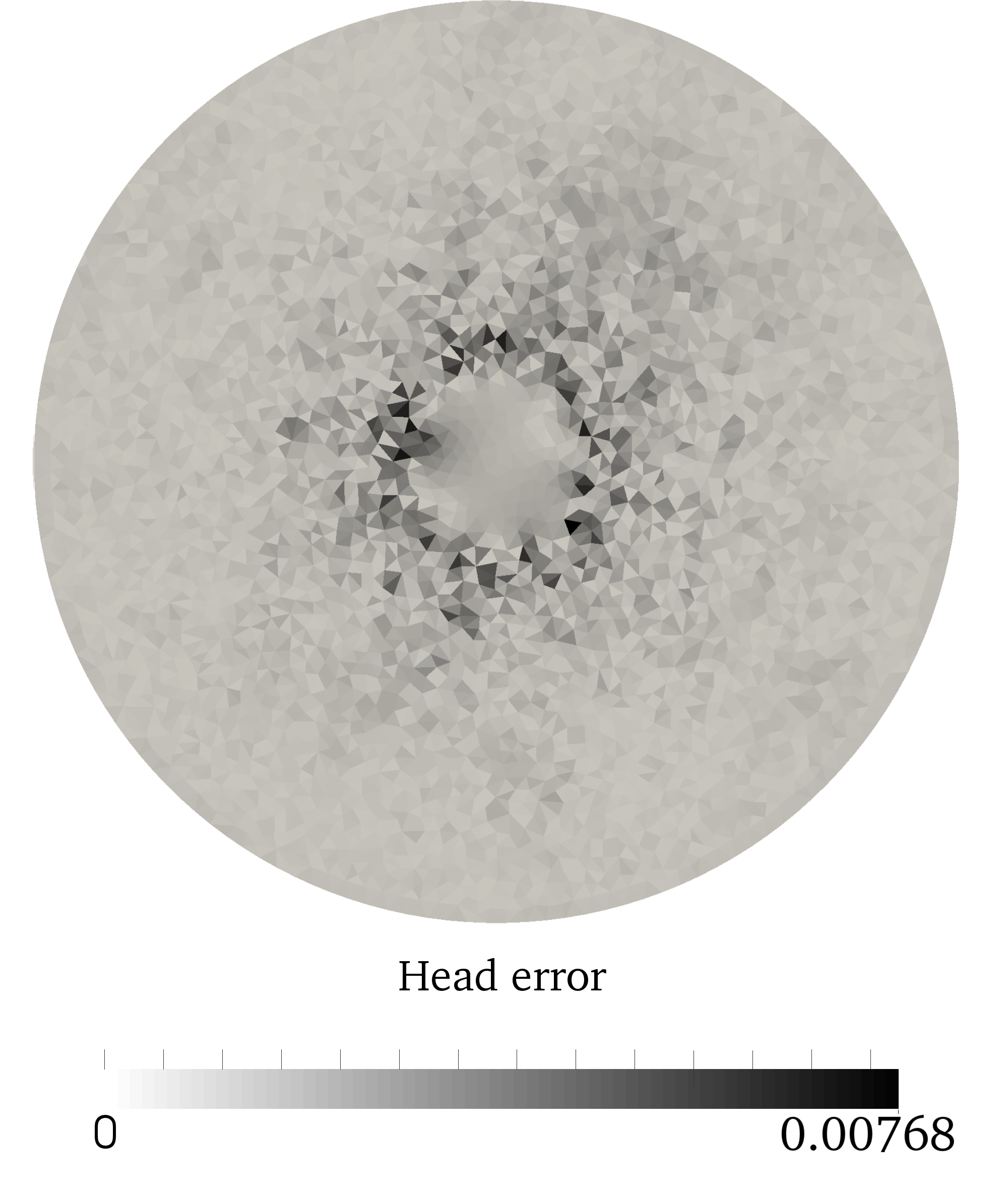}\\
   \end{tabular}
\caption{Absolute hydraulic head error using mesh $P=4\sqrt{2}$ in Example \ref{example1} with the uncorrected scheme (left), WFC scheme (middle), and NWC scheme (right) for the well radius $r=0.05$.}
\label{headError1}       
\end{figure}

From analytical solution (\ref{pr1}) we can see that the gradient of hydraulic head changes more sharply if the well radius is smaller. Therefore, we can expect better results with the uncorrected scheme if the well is larger. 

It follows from equations (\ref{wellFaceFluxCorr}) and (\ref{diffBun}) that the distance between cell centroids and the well center must not be less than the well radius. This requirement limits how much we can refine the grids. This should not present a problem in real-world applications, but in order to be able to perform the refinement tests, we triangulate the ring domain and use the inner circle of the ring as the well cell.

\begin{table}[!htbp]
\caption{Errors of the uncorrected scheme for the well radius $r=1$ in Example \ref{example1}.}
\label{table12}   
\begin{center}
\begin{tabular}{c c c c c c c}
\hline
$P$ & 32$\sqrt{2}$ & 16$\sqrt{2}$ & 8$\sqrt{2}$ & 4$\sqrt{2}$ & 2$\sqrt{2}$ & $\sqrt{2}$ \\
\hline
$\epsilon_{2}^{h}$ & 8.32e-02 & 5.22e-02 & 3.36e-02 & 1.83e-02 & 8.37e-03 & 2.59e-03 \\
$\epsilon_{\max}^{h}$  & 2.22e-01 & 1.85e-01 & 1.48e-01 & 1.05e-01 & 5.99e-02 & 3.42e-02\\
$\epsilon_{Q}$ & 1.45e+00 & 8.85e-01 & 5.68e-01 & 3.10e-01 & 1.43e-01 & 4.43e-02 \\
$n_{\text{Pic}}$ & 9 & 11 & 11 & 11 & 12 & 11\\
\\
\noalign{\smallskip}\hline
\end{tabular}
\end{center}
\end{table}

\begin{table}[!htbp]
\caption{Errors of the uncorrected scheme for the well radius $r=50$ in Example \ref{example1}.}
\label{table13}   
\begin{center}
\begin{tabular}{c c c c c c c}
\hline
$P$ & 32$\sqrt{2}$ & 16$\sqrt{2}$ & 8$\sqrt{2}$ & 4$\sqrt{2}$ & 2$\sqrt{2}$ & $\sqrt{2}$ \\
\hline
$\epsilon_{2}^{h}$ & 1.74e-03 & 5.08e-04 & 1.42e-04 & 3.49e-05 & 9.15e-06 & 2.41e-06 \\
$\epsilon_{\max}^{h}$  & 3.81e-03 & 1.61e-03 & 4.94e-04 & 1.27e-04 & 4.76e-05 & 9.75e-06\\
$\epsilon_{Q}$ & -1.83e-03 & -8.60e-04 & -3.17e-04 & -7.59e-05 & -2.03e-05 & -5.85e-06 \\
$n_{\text{Pic}}$ & 11 & 11 & 11 & 11 & 10 & 9\\
\\
\noalign{\smallskip}\hline
\end{tabular}
\end{center}
\end{table}

Results presented in Tables \ref{table12} and \ref{table13} show that the uncorrected scheme approaches first-order accuracy when the well radius is $r=1$ and second-order accuracy when the well radius is $r=50$.

\end{example}
\begin{example}
\label{example2}
Here we consider a rectangular reservoir with corners $(\pm300,\pm150)$ and with hydraulic conductivity $K=0.0001$. Two wells with radii $r_{\text{l}}$ and $r_{\text{r}}$ are specified at $(-150,0)$ and $(150,0)$, respectively. 

An analytical solution is obtained by superposing two solutions of form (\ref{pr1}):
\begin{equation}
h({\bf x}) = \frac{h_{\text{l}} \ln \frac{R_{\text{l}}}{\rho_{\text{l}}}+h_{R_{\text{l}}}\ln\frac{\rho_{\text{l}}}{r_{\text{l}}}}{\ln \frac{R_{\text{l}}}{r_{\text{l}}}}+\frac{h_{\text{r}}\ln \frac{R_{\text{r}}}{\rho_{\text{r}}}+h_{R_{\text{r}}}\ln \frac{\rho_{\text{r}}}{r_{\text{r}}}}{\ln \frac{R_{\text{r}}}{r_{\text{r}}}},
\label{pr2}
\end{equation}
where the distances from the left and the right well are denoted by $\rho_{l}$ and $\rho_{r}$, respectively. We take $h_{\text{l}}=5$, $h_{\text{r}}=10$, $h_{R_{\text{l}}}=h_{R_{\text{r}}}=20$, $r_{\text{l}}=0.5$, $r_{\text{r}}=0.6$ and $R_{\text{l}}=R_{\text{r}}=1200$. Note that in this case $h_{R_{\text{l}}}$, $h_{R_{\text{r}}}$, $h_{\text{l}}$, $h_{\text{r}}$ are just formal parameters. In engineering practice these parameters are obtained when one well is turned off.This is a slightly different approach to finding analytical solution for two wells than in \cite{Hai95}. 

Transfer coefficient $\Psi$ is set for each well face separately, so that (\ref{colmation}) and (\ref{pr2}) give level $23$ in the left well and $27$ in the right well. On the outer boundary of the domain we prescribe the exact hydraulic head obtained from equation (\ref{pr2}).
\begin{table}[!htbp]
\caption{Errors in Example \ref{example2}.}
\label{table2}       
\begin{center}
\begin{tabular}{c c c c c c c}
\hline
$P$ & 64 & 32 & 16 & 8 & 4 & 2 \\
\hline
\multicolumn{7}{l}{Uncorrected scheme}\\
$\epsilon_{2}$ & 1.22e-01 & 7.81e-02 & 4.39e-02 & 3.03e-02 & 1.57e-02 & 9.10e-03 \\
$\epsilon_{\max}$  & 2.65e-01 & 2.41e-01 & 1.82e-01 & 1.56e-01 & 9.88e-02 & 7.35e-02\\
$\epsilon_{Q_{\text{l}}}$ & 3.56e-00 & 2.18e-00 & 1.15e-00 & 7.83e-01 & 4.05e-01 & 2.50e-01 \\
$\epsilon_{Q_{\text{r}}}$ & 2.83e-00 & 1.66e-00 & 9.97e-01 & 7.21e-01 & 3.78e-01 & 1.91e-01 \\
$n_{\text{Pic}}$ & 7 & 11 & 10 & 11 & 11 & 13 \\
\\
\multicolumn{7}{l}{WFC scheme}\\
$\epsilon_{2}$ & 1.98e-03 & 1.42e-03 & 7.60e-04 & 7.33e-04 & 7.03e-04 & 6.89e-04 \\
$\epsilon_{\max}$ & 6.71e-02 & 1.03e-02 & 8.11e-03 & 9.08e-03 & 7.72e-03 & 7.70e-03\\
$\epsilon_{Q_{\text{l}}}$  & 1.74e-02 & 1.88e-02 & 1.71e-02 & 1.74e-02 & 1.75e-02 & 1.71e-02 \\
$\epsilon_{Q_{\text{r}}}$ & 1.54e-02 & 1.64e-02 & 1.42e-02 & 1.69e-02 & 2.18e-02 & 2.56e-02 \\
$n_{\text{Pic}}$ & 8 & 11 & 10 & 11 & 11 & 13 \\
\\
\multicolumn{7}{l}{NWC scheme}\\
$\epsilon_{2}$ & 9.19e-04 & 1.64e-04 & 3.20e-05 & 8.16e-06 & 1.97e-06 & 5.52e-07\\
$\epsilon_{\max}$ & 2.28e-03 & 5.17e-04 & 1.80e-04 & 4.84e-05 & 1.14e-05 & 3.49e-06\\
$\epsilon_{Q_{\text{l}}}$ & 8.11e-03 & 6.16e-04 & 2.20e-05 & -1.13e-05 & 6.55e-06 & 3.44e-06\\
$\epsilon_{Q_{\text{r}}}$ & 7.88e-03 & 2.35e-04 & -4.36e-05 & -5.24e-05 & -4.70e-06 & 1.37e-06 \\
$n_{\text{Pic}}$ & 8 & 11 & 10 & 14 & 14 & 13 \\
\noalign{\smallskip}\hline
\end{tabular}
\end{center}
\end{table}

\begin{figure}[!htbp]
\centering
   \begin{tabular}{@{}cc@{}}
   \includegraphics[width=80mm]{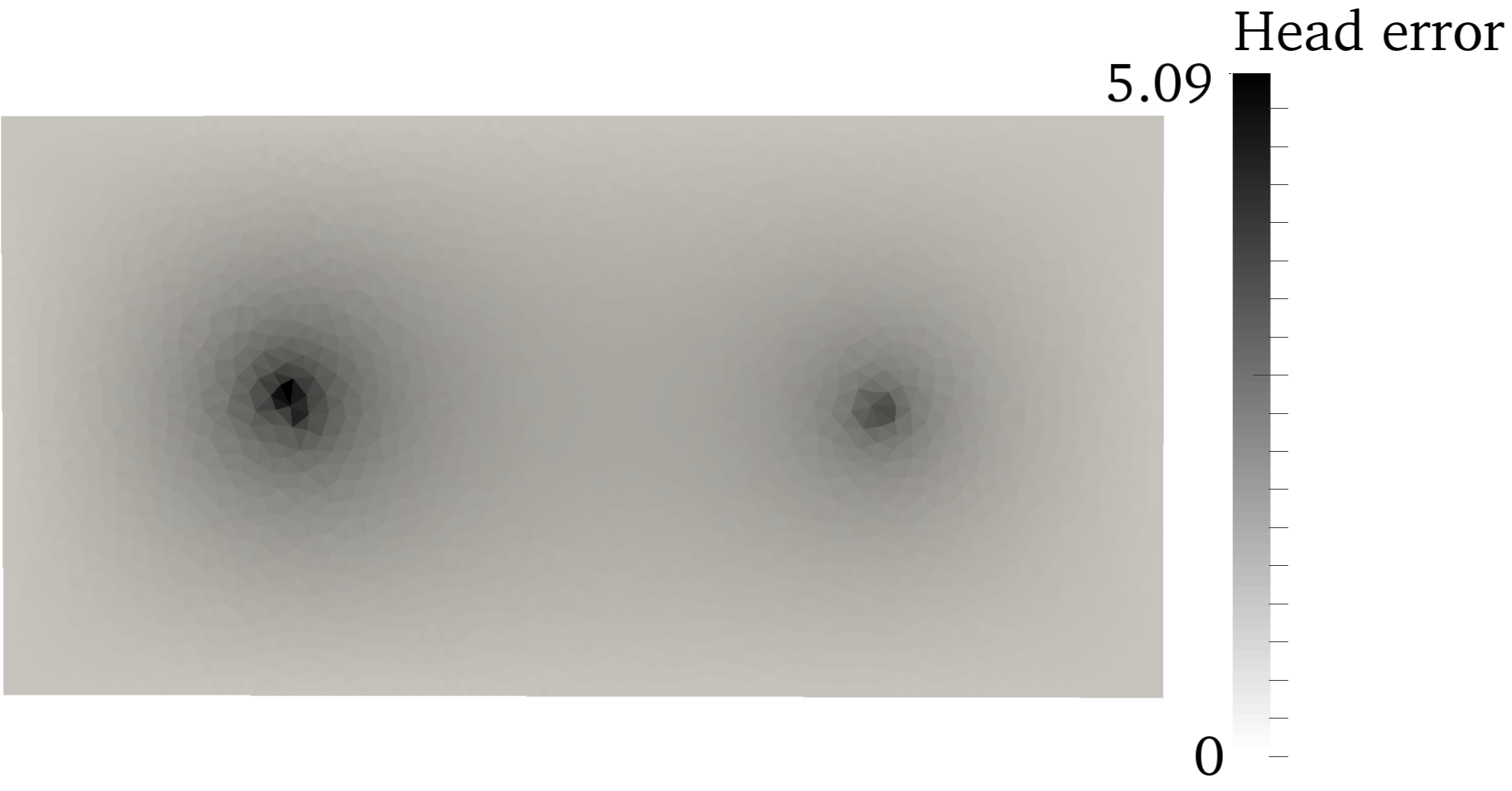}&
   \includegraphics[width=80mm]{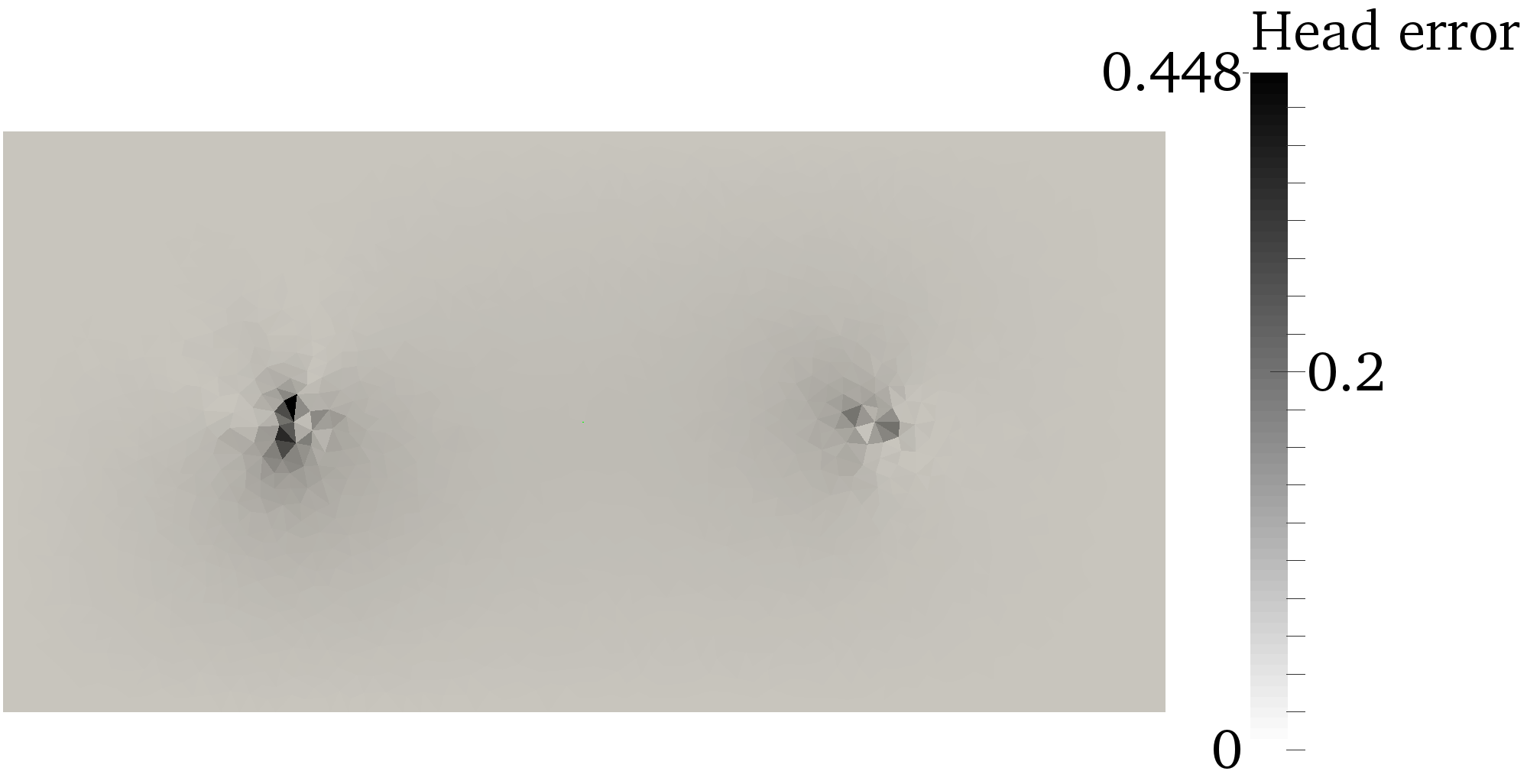}\\
   \multicolumn{2}{c}{\includegraphics[width=80mm]{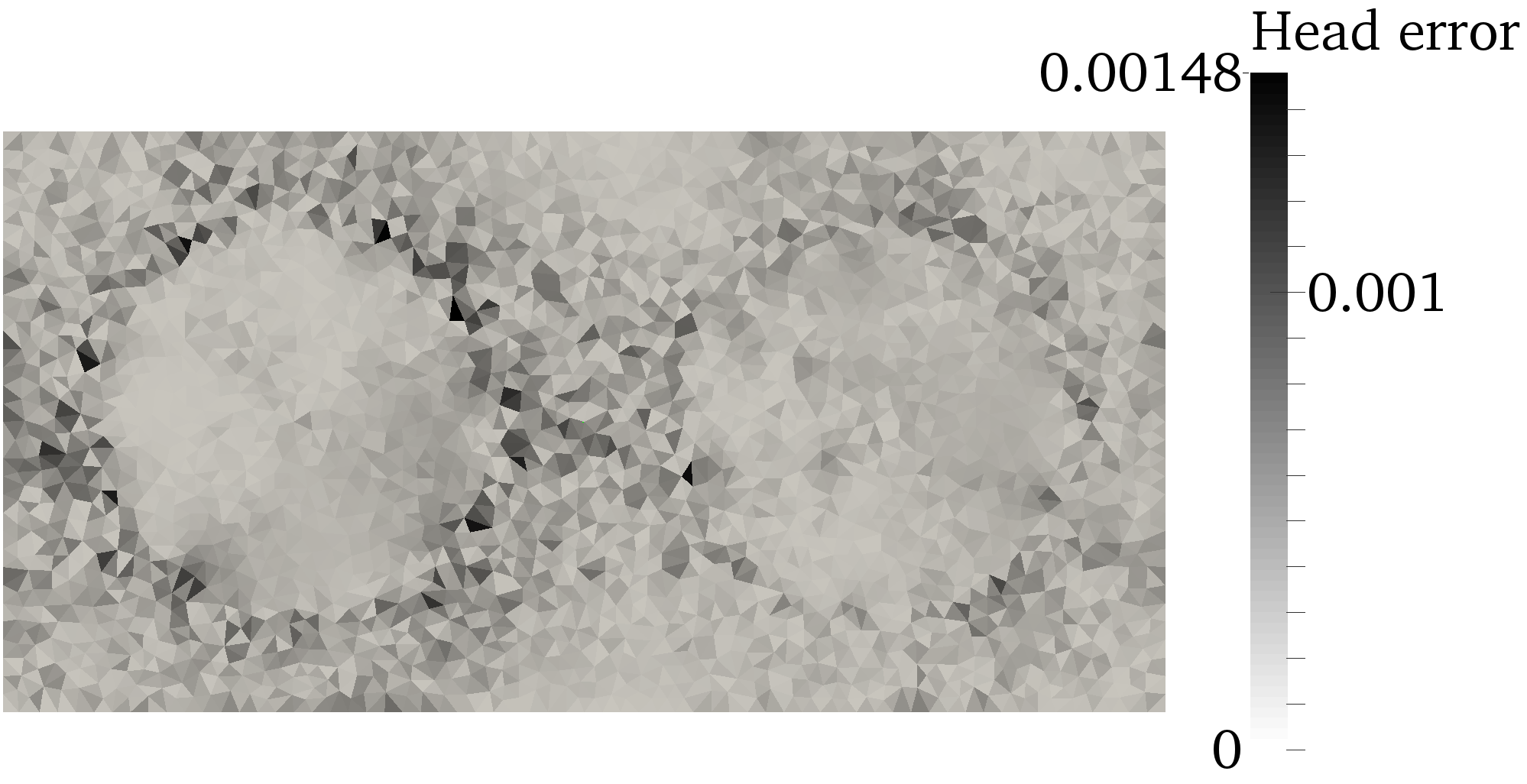}}
   \end{tabular}
\caption{Absolute hydraulic head error using mesh $P=8$ in Example \ref{example2} with the uncorrected scheme (top left), WFC scheme (top right), and the NWC scheme (bottom).}
\label{headError2}       
\end{figure}

As in the previous example, the uncorrected scheme is not second-order accurate and the well flow rates are very inaccurate (Table \ref{table2}). The total well flux error is much smaller with the WFC scheme, but the scheme is not convergant.  
The results for the NWC scheme are obtained using a circular near-well region with radius $100$. These results show that the NWC scheme is second-order accurate. 

\end{example}

\begin{example}
\label{example21}
In this example we examine the same domain with two wells as in the previous example. In the left well with radius $r_{\text{l}}=0.5$ we prescribe hydraulic head $h_{\text{l}}=0$ and in the right well with radius $r_{\text{r}}=0.6$ we prescribe hydraulic head $h_{\text{r}}=1$. A no-flow condition ($g_{\text{N}}=0$) is set at the outer boundaries.

The maximum principle guarantees that the exact solution is between 0 and 1.
It is well known \cite{Dan09,Vid11} that non-linear two-point flux approximation preserves positivity of the obtained discrete solution, but violates the upper limit. The aim of this example is to show that NWC and WFC inherits this property, i.e. the discret solution preserves the solution positivity.

\begin{table}[!htbp]
\caption{Minimal value of the hydraulic head in Example \ref{example21}.}
\label{table21}   
\begin{center}
\begin{tabular}{c c c c c }
\hline
$P$ & 64 & 32 & 16 & 8  \\
\hline
Uncorrected scheme & 5.88e-05 & 4.42e-05 & 3.51e-05 & 3.12e-05  \\
WFC scheme & 2.11e-07 & 2.10e-07 & 2.09e-07 & 2.09e-07  \\
NWC scheme & 2.08e-07 & 2.07e-07 & 2.07e-07 & 2.06e-07  \\
\noalign{\smallskip}\hline
\end{tabular}
\end{center}
\end{table}

Results in the Table \ref{table21} shows that obtained discrete solution preserves positivity.

\end{example}

\begin{example}
\label{example3}
The domain is a box with corners $(\pm 100,\pm 50, \pm 50)$. It contains two straight wells, one horizontal from $(-50,-50,0)$ to $(-50, 50,0)$ and one vertical from $(50,0,-50)$ to $(50,0,50)$. 

An analytical solution is again obtained by superposition and is given by (\ref{pr2}). 
Distances $\rho_{\text{l}}$ and $\rho_{\text{r}}$ are calculated as
\begin{equation}
\rho_{\text{l}}=\sqrt{(x-x_{\text{l}})^{2}+(z-z_{\text{l}})^{2}},
\quad \rho_{\text{r}}=\sqrt{(x-x_{\text{r}})^{2}+(y-y_{\text{r}})^{2}},
\end{equation}
where $x_{\text{l}}=-50$, $z_{\text{l}}=0$, $x_{\text{r}}=50$ and $y_{\text{r}}=0$. 

In this example we take $R_{\text{l}}=R_{\text{r}}=1000$, $h_{R_{\text{l}}}=50$, $h_{R_{\text{r}}}=53$, $r_{\text{l}}=0.1$, $r_{\text{r}}=0.15$, $h_{\text{l}}=40$ and $h_{\text{r}}=45$.

The transfer coefficient in each well face is chosen according to the Hagen-Poiseuille law so that the head in the horizontal well pump is $90$ and the head in the vertical well pump is $92$. The pumps are located at $(-50,-50,0)$ and $(50,0,-50)$ for the horizontal and vertical wells, respectively. Hydraulic head isosurfaces are shown in Fig. \ref{3primer} on the left and the mesh (for $P=8$) is shown on the right.

The errors of the uncorrected, WFC, and NWC schemes are shown in Table \ref{table3}. A near-well region of radius $30$ is used.

As in the previous examples, only the NWC scheme is second-order accurate.

\begin{figure}[!htbp]
\centering
  \includegraphics[width=140mm]{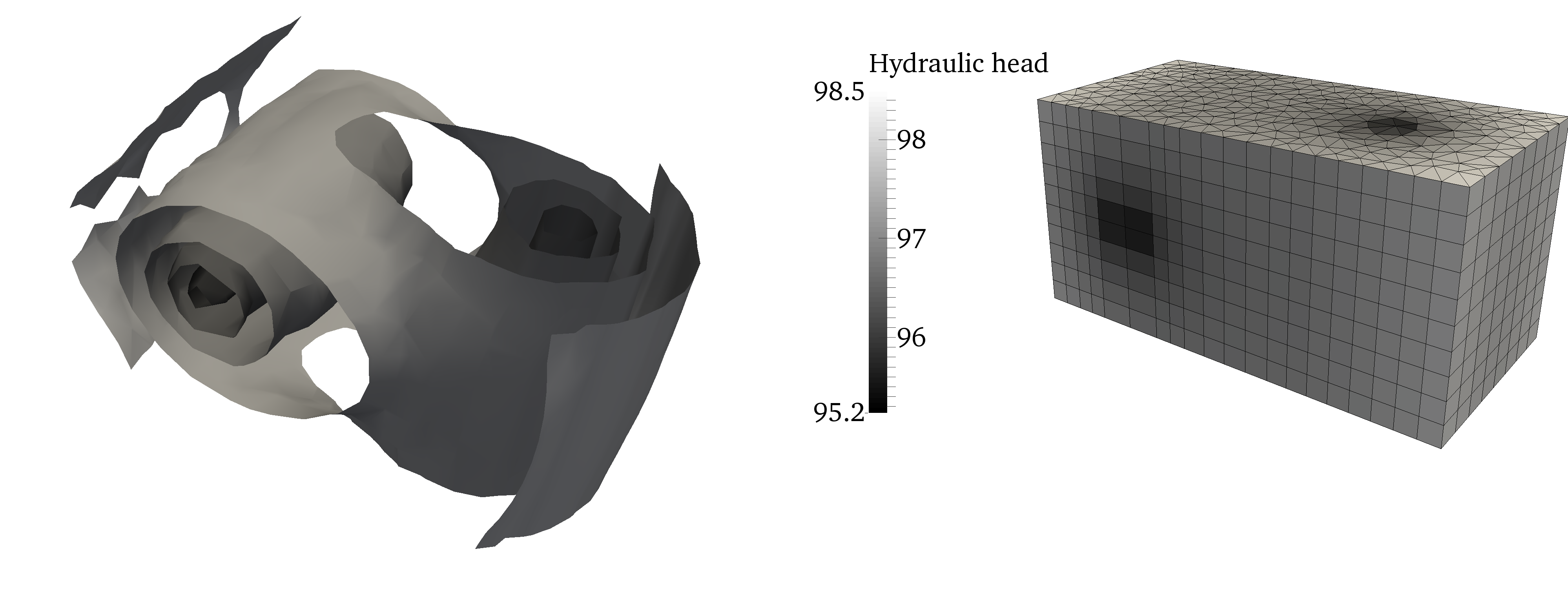}
\caption{Hydraulic head isosurfaces (left) and the mesh (right) in Example \ref{example3}.}
\label{3primer}      
\end{figure}

\begin{table}[!hbp]
\caption{Errors in Example \ref{example3}.}
\label{table3}       
\begin{center}
\begin{tabular}{c c c c c}
\hline
$P$ & 16 & 8  & 4 & 2 \\
\hline
\multicolumn{5}{l}{Uncorrected scheme}\\
$\epsilon_{2}$ & 3.84e-03 & 3.63e-03 & 2.78e-03 & 1.70e-03  \\
$\epsilon_{\max}$  & 1.82e-02 & 1.83e-02 & 2.01e-02 & 1.68e-02 \\
$\epsilon_{Q_{\text{l}}}$ & -8.54e-01 & -8.22e-01 & -6.75e-01 & -4.52e-01  \\
$\epsilon_{Q_{\text{r}}}$ & -8.33e-01 & -7.30e-01 & -4.84e-01 & -2.20e-01  \\
$n_{\text{Pic}}$ & 23 & 14 & 12 & 11 \\
\\
\multicolumn{5}{l}{WFC scheme}\\
$\epsilon_{2}$ & 4.86e-04 & 1.85e-04 & 9.07e-05 & 6.13e-05 \\
$\epsilon_{\max}$ & 3.36e-03 & 2.41e-03 & 2.02e-03 & 1.25e-03 \\
$\epsilon_{Q_{\text{l}}}$  & 2.01e-02 & 1.32e-02 & 9.19e-03 & -9.43e-03 \\
$\epsilon_{Q_{\text{r}}}$ & 1.99e-02 & 1.92e-02 & 1.81e-02 & 1.54e-02 \\
$n_{\text{Pic}}$ & 17 & 15 & 14 & 14 \\
\\
\multicolumn{5}{l}{NWC scheme}\\
$\epsilon_{2}$ & 6.35e-05 & 1.72e-05 & 4.49e-06 & 1.25e-06 \\
$\epsilon_{\max}$ & 2.22e-04 & 5.54e-05 & 2.70e-05 & 6.26e-06 \\
$\epsilon_{Q_{\text{l}}}$ & 1.81e-03 & 5.99e-04 & 5.55e-05 & -1.77e-05 \\
$\epsilon_{Q_{\text{r}}}$ & 2.16e-04 & 4.68e-04 & 1.82e-05 & 1.01e-05  \\
$n_{\text{Pic}}$ & 14 & 17 & 19 & 23 \\
\noalign{\smallskip}\hline
\end{tabular}
\end{center}
\end{table}

\end{example}

\begin{example}
\label{example5}
Circular domain $R=10$ with a well of radius $r=1$ in the center $(0,0)$ is considered. Hydraulic conductivity
\begin{equation}
K=-4.5 \cdot 10^{-6}\cdot (x+10)+10^{-4}
\end{equation}
varies between $10^{-4}$ and $10^{-5}$.
We specify the hydraulic head $h_{\text{w}}=55$ in the well and set $h_{R}=75$ at $\rho=R$. In this example there is no colmation.

We use meshes with parameter $P=1$, $1/2$, $1/4$, $1/8$, $1/16$, and $1/32$. Since an analytical solution is not available, we compare these results to the solution obtained with the uncorrected method on a mesh with parameter $P=1/64$. In practice, meshes as fine as these can rarely be used, but we give this example in order to demonstrate that even the uncorrected scheme becomes second-order accurate on fine meshes, and to verify the WFC and NWC schemes in the inhomogeneous case. Of course in this way we can only demonstrate that the approximate solutions converge to some limit at a certain rate and not that this limit is the actual solution. However this has been demonstrated for homogeneous $\mathbb{K}$ in example \ref{example1}. A near-well region of radius 2 is used for the NWC scheme. 

As in Example \ref{example1} for $r=50$, we triangulate the ring domain and use the inner circle of the ring as the well cell.

The errors of the uncorrected scheme are shown in Table \ref{table5}. The order of accuracy is less than two on coarse meshes, but on finer meshes this scheme is second-order accurate.  

\begin{table}[!htbp]
\caption{Errors of the uncorrected scheme in Example \ref{example5}.}
\label{table5}       
\begin{center}
\begin{tabular}{c c c c c c c}
\hline
$P$ & 1 & 1/2  & 1/4 & 1/8 & 1/16 & 1/32\\
\hline
$\epsilon_{2}$ &  5.93e-04 & 2.01e-04 & 9.95e-05 & 3.01e-05 & 7.31e-06 & 1.52e-06 \\
$\epsilon_{\max}$  & 2.40e-03 & 1.58e-03 & 6.95e-04 & 2.06e-04 & 5.46e-05 & 1.55e-05 \\
$\epsilon_{Q}$ &  -7.61e-03 & -2.27e-03 & -1.18e-03 & -3.70e-04  & -8.78e-05 & -1.87e-05\\
$n_{\text{Pic}}$ & 12 & 12 & 14 & 15 & 15 & 15\\
\noalign{\smallskip}\hline
\end{tabular}
\end{center}
\end{table}

Table \ref{table51} shows that the obtained errors with WFC scheme are smaller than with the uncorrected scheme. The results also show that the hydraulic head obtained with the NWC scheme is second-order accurate. 

\begin{table}[!htbp]
\caption{Errors of the WFC and NWC schemes in Example \ref{example5}.}
\label{table51}       
\begin{center}
\begin{tabular}{c c c c c}
\hline
$P$ & 1 & 1/2  & 1/4 & 1/8 \\
\hline

\multicolumn{5}{l}{WFC scheme}\\
$\epsilon_{2}$ & 4.51e-04 & 6.72e-05 & 1.51-05 & 4.30e-06 \\
$\epsilon_{\max}$ & 1.95e-03 & 9.33e-04 & 3.11e-04 & 8.17e-05 \\
$\epsilon_{Q}$  & -5.78e-03 & -2.46e-04 & -8.44e-05 & -1.58e-05 \\
$n_{\text{Pic}}$ & 12 & 12 & 14 & 14 \\
\\
\multicolumn{5}{l}{NWC scheme}\\
$\epsilon_{2}$ & 9.16e-05 & 2.36e-05 & 6.56e-06 & 2.03e-06 \\
$\epsilon_{\max}$ & 4.87e-04 & 1.61e-05 & 6.53e-05 & 1.88e-05 \\
$\epsilon_{Q}$ & -2.85e-04 & -6.65e-05 & 2.09e-05 & 4.89e-06 \\
$n_{\text{Pic}}$ & 13 & 13 & 15 & 20 \\
\noalign{\smallskip}\hline
\end{tabular}
\end{center}
\end{table}

\end{example}

\begin{example}
\label{example4}
We consider the same domain as in Example \ref{example1} with the well radius $r=0.05$, and with a heterogeneous hydraulic conductivity 
\begin{equation}
K=\left( \sin \frac{\pi x}{300} \cdot \sin\frac{\pi y}{300}+1\right)\cdot10^{-4}.
\end{equation}
We take $h_{R}=100$, $h_{\text{w}}=60$, and assume that $h_{\text{w}}=h_{r}$. 

The analytical solution to this problem is not known, therefore we compare the obtained results with the solution computed using the NWC scheme and mesh with $P=\sqrt{2}/2$. In Table \ref{table41} we show the scaled norm of the differences between the solutions obtained with the NWC scheme using near-well zones of radius 20 and 50. The norms were computed in the same way as the errors in (\ref{l2Error}) and (\ref{maxError}). These solutions approach each other quadratically, which tells us that we can compute the referent solution on the finest grid using any near-well zone radius. We use a near-well zone with radius 50.

\begin{table}[!ht]
\caption{Differences of solutions obtained using the NWC scheme with the near-well zone of radius $20$ and $50$ in Example \ref{example4}.}
\label{table41}       
\begin{center}
\begin{tabular}{c c c c c c c}
\hline
$P$ & $16\sqrt{2}$  & $8\sqrt{2}$ & $4\sqrt{2}$ & $2\sqrt{2}$ & $\sqrt{2}$ & $\sqrt{2}/2$\\
\hline
2-norm & 3.90e-04 & 1.23e-04 & 2.06e-05 & 4.99e-06 & 1.15e-06 & 2.96e-07\\
max norm  & 2.47e-03 & 1.59e-03 & 3.47e-04 & 1.15e-04 & 2.86e-05 & 1.19e-05\\
\noalign{\smallskip}\hline
\end{tabular}
\end{center}
\end{table}

The norms of differences from the referent solution are presented in Table \ref{table4}. As in the previous examples, the NWC scheme appears to be second-order accurate, and with the WFC scheme the accuracy is greatly improved but the scheme is still inconsistent.

\begin{table}[!ht]
\caption{Errors in Example \ref{example4}.}
\label{table4}       
\begin{center}
\begin{tabular}{c c c c c}
\hline
$P$ & $32\sqrt{2}$ & $16\sqrt{2}$  & $8\sqrt{2}$ & $4\sqrt{2}$ \\
\hline
\multicolumn{5}{l}{Uncorrected scheme}\\
$\epsilon_{2}$ & 1.48e-01 & 1.05e-01 & 6.57e-02 & 4.87e-02  \\
$\epsilon_{\max}$  & 3.78e-01 & 3.35e-01 & 2.81e-01 & 2.47e-01 \\
$\epsilon_{Q}$ & 3.54e-00 & 2.53e-00 & 1.62e-00 & 1.21e-00  \\
$n_{\text{Pic}}$ & $8$ & $10$ & $11$ & $11$ \\
\\
\multicolumn{5}{l}{WFC scheme}\\
$\epsilon_{2}$ & 2.01e-03 & 8.46e-04 & 6.25-04 & 5.24e-04 \\
$\epsilon_{\max}$ & 1.13e-02 & 7.61e-03 & 9.15e-03 & 6.05e-03 \\
$\epsilon_{Q}$ & 1.55e-02 & 1.42e-02 & 1.40e-02 & 1.32e-02 \\
$n_{\text{Pic}}$ & $8$ & $10$ & $11$ & $11$ \\
\\
\multicolumn{5}{l}{NWC scheme}\\
$\epsilon_{2}$ & 1.64e-03 & 3.39e-04 & 7.17e-05 & 1.92e-05 \\
$\epsilon_{\max}$ & 1.12e-02 & 1.42e-03 & 5.94e-04 & 2.82e-04 \\
$\epsilon_{Q}$ & 4.41e-03 & 1.15e-03 & -3.20e-04 & 3.30e-05  \\
$n_{\text{Pic}}$ & $8$ & $10$ & $11$ & $11$ \\
\noalign{\smallskip}\hline
\end{tabular}
\end{center}
\end{table}

\end{example}

\begin{example}
\label{example6}
We consider a discontinuous circular reservoir with a well in the center. Hydraulic conductivity is 
\begin{equation}
K=\left\{\begin{array}{l}
K_{1} \quad\text{if } y<0,\\
K_{2} \quad\text{otherwise},
\end{array}
\right.
\quad
K_{1}=10^{-3},
\quad
K_{2}=10^{-6}.
\end{equation}

The exact hydraulic head is given by (\ref{pr1}), while the exact flow rate is 
\begin{equation}
Q=\pi \left(K_{1} + K_{2}\right) \frac{h_{R}-h_{r}}{\ln\frac{R}{r}}.
\end{equation}

We take $r=0.05$, $R=200$, $h_{\text{w}}=55$, and $h_{R}=100$. Transfer coefficient $\Psi$ is set for each well face separately so that the hydraulic head at the well wall is $h_{r}=60$. 

The errors of the uncorrected, WFC, and NWC schemes are shown in Table \ref{table6}. A circular near-well region of radius $50$ is used. As in the homogeneous case, the WFC scheme gives improved results in comparison to the uncorrected scheme, but only the NWC scheme is second-order accurate.

\begin{table}[!htbp]
\caption{Errors in Example \ref{example6}.}
\label{table6}   
\begin{center}
\begin{tabular}{c c c c c c}
\hline
$\mathfrak{h}$ & 16$\sqrt{2}$ & 8$\sqrt{2}$ & 4$\sqrt{2}$ & 2$\sqrt{2}$ & $\sqrt{2}$ \\
\hline
\multicolumn{6}{l}{Uncorrected scheme}\\
$\epsilon_{2}^{h}$ & 6.10e-02 & 4.34e-02 & 3.92e-02 & 2.36e-02 & 2.03e-02  \\
$\epsilon_{\max}^{h}$  & 2.23e-01 & 2.01e-01 & 1.95e-01 & 1.60e-01 & 1.43e-01 \\
$\epsilon_{Q}$ & 1.65e+00 & 1.14e+00 & 1.05e+00 & 6.25e-01 & 5.45e-01  \\
$n_{P}$ & 12 & 12 & 12 & 13 & 13 \\
\\
\multicolumn{6}{l}{WFC scheme}\\
$\epsilon_{2}^{h}$ & 1.15e-03 & 5.71e-04 & 4.55e-04 & 3.76e-04 & 4.23e-04 \\
$\epsilon_{\max}^{h}$ & 1.26e-02 & 7.83e-03 & 1.41e-02 & 8.14e-03 & 1.56e-02 \\
$\epsilon_{Q}$  & 1.72e-02 & 1.17e-02 & 1.18e-02 & 7.06e-03 & 1.19e-02 \\
$n_{P}$ & 12 & 12 & 12 & 13 & 13 \\
\\
\multicolumn{6}{l}{NWC scheme}\\
$\epsilon_{2}^{h}$ & 1.76e-04 & 6.03e-05 & 1.32e-05 & 3.01e-06 & 7.00e-07 \\
$\epsilon_{\max}^{h}$ & 6.75e-04 & 4.69e-04 & 9.05e-05 & 2.55e-05 & 5.97e-06 \\
$\epsilon_{Q}$ & 1.25e-04 & -3.36e-05 & 7.17e-06 & -3.14e-06 & -3.39e-06 \\
$n_{P}$ & 12 & 12 & 12 & 14 & 14 \\
\noalign{\smallskip}\hline
\end{tabular}
\end{center}
\end{table}

\end{example}

\section{Conclusion}

Discretization schemes based on linear approximations produce very inaccurate results on coarse grids if a well is present. On very fine meshes, even this type of scheme can produce a second-order accurate solution as shown in Example \ref{example5}. However, such fine meshes can rarely be used in practice. The uncorrected scheme canalso achieve second-order accuracy on locally refined meshes, if the mesh size in the well viscinity is smallerthan the well radius, but this comes at a high computational cost.

We have developed two schemes for the discretization of near-well fluxes.  

The first scheme (WFC scheme, Section \ref{Correction}) reduces the hydraulic head and flowrate errors, but it is not convergent unless the grids are very fine.

Numerical examples show that the second scheme (NWC scheme, Section \ref{nearWellScheme}) gives at least a first-order accurate total well flux and a second-order accurate hydraulic head without near-well local mesh refinement.

Both schemes were developed for the case of an isotropic hydraulic conductivity. An extension of these schemes to the anisotropic homogeneous case was presented in \cite{Dot14}.

The one-side flux approximation (\ref{fluxApp}) can also be used in scheme \cite{Dro11}, that preserves minimum and maximum principles. This has been implemented in WODA, an open-source groundwater solver \cite{WODA}. Preliminary results indicate that such a scheme is second-order accurate in the well vicinity and preserves the minimum and maximum principles.

\appendix
\section{Lambert cylindrical equal-area projection}
\label{AppendixB}
\begin{figure}[!ht]
\begin{center}
  \includegraphics[width=65mm]{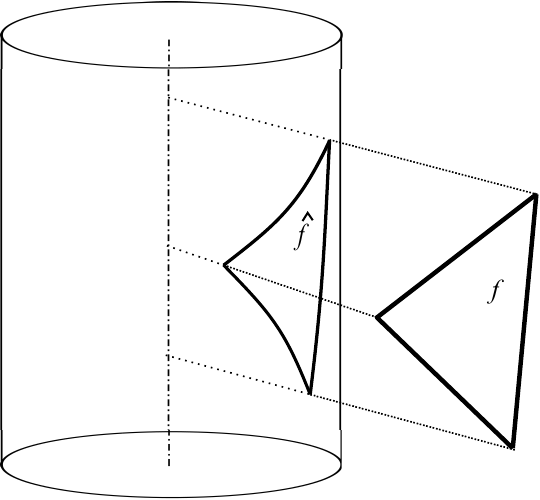}
\caption{Projection of triangular face onto cylinder.}
\label{proj} 
\end{center}
\end{figure}
Projection of ${\bf x}$ onto a cylinder is defined by 
\begin{equation}
\text{Pr}({\bf x})={\bf x}_{p}+r\frac{({\bf x}_{i}-{\bf x}_{p})}{\|{\bf x}_{i}-{\bf x}_{p}\|},
\end{equation}
where ${\bf x}_{p}$ is the orthogonal projection of ${\bf x}$ onto the cylinder axis and $r$ is the cylinder radius. 

The projection of a straight line is generally not a second-order curve (Fig. \ref{proj}). Numerical integration is used to calculate the area of Pr$(f)$ in Section \ref{3D}. The results presented in this paper were obtained using the 6\textsuperscript{th} order Gauss-Legendre integration formula. In our case this formula was accurate enough to calculate the integrals with machine precision.

\section*{Acknowledgments}
The research leading to these results has received funding from the Serbian Ministry of Education, Science and Technological Development under a project titled: Methodology for Assessment, Design and Maintenance of Groundwater Source in Alluvials Depending on Aerobic Level, No. TR37014.

\section*{\refname}
\bibliographystyle{elsarticle-num}  
\bibliography{biblioT}

\begin{thebibliography}{10}
\expandafter\ifx\csname url\endcsname\relax
  \def\url#1{\texttt{#1}}\fi
\expandafter\ifx\csname urlprefix\endcsname\relax\def\urlprefix{URL }\fi
\expandafter\ifx\csname href\endcsname\relax
  \def\href#1#2{#2} \def\path#1{#1}\fi

\bibitem{Dim11a}
M.~Dimki\'c, M.~Pu\v{s}i\'c, D.~Vidovi\'c, N.~Filipovi\'c, V.~Isailovi\'c,
  B.~Majki\'c, Numerical model assessment of radial-well aging, ASCE's Journal
  of computing in civil engineering 25~(1) (2011) 43--49.

\bibitem{Dim14}
M.~Dimki\'c, M.~Pu\v{s}i\'c, Correlation between entrance velocities, increase
  in local hydraulic resistances and redox potential of alluvial groundwater
  sources, Water Research and Managment 4~(4) (2014) 3--33.

\bibitem{Dan09}
A.~Danilov, Y.~Vassilevski, A monotone nonlinear finite volume method for
  diffusion equations on conformal polyhedral meshes, Russ. J. Numer. Anal.
  Math. Modelling 24~(3) (2009) 207--227.

\bibitem{Lep05}
C.~{Le Potier}, Sch\'ema volumes finis monotone pour des op\'erateurs de
  diffusions fortement anisotropes sur des maillages de triangle non
  structur\'es, C.R. Math. Acad. Sci. Paris 341 (2005) 787--792.

\bibitem{Lip07}
K.~Lipnikov, M.~Shashkov, D.~Svyatskiy, Y.~Vassilevski, Monotone finite volume
  schemes for diffusion equations on unstructured triangular and shape-regular
  polygonal meshes, J. Comp. Phys. 227~(1) (2007) 492--512.

\bibitem{Vas08}
Y.~Vassilevski, I.~Kapyrin, Two splitting schemes for nonstationary
  convection-diffusion problems on tetrahedral meshes, Comput. Math. Math.
  Phys. 48~(8) (2008) 1349--1366.

\bibitem{Vid11}
D.~Vidovi\'c, M.~Dimki\'c, M.~Pu\v{s}i\'c, Accelerated non-linear finite volume
  method for diffusion, J. Comp. Phys. 230~(7) (2011) 2722--2735.

\bibitem{Vid13}
D.~Vidovi\'c, M.~Dotli\'c, M.~Dimki\'c, M.~Pu\v{s}i\'c, B.~Pokorni, Convex
  combinations for diffusion schemes, J. Comp. Phys. 246 (2013) 11--27.

\bibitem{Vid14}
D.~Vidovi\'c, M.~Dotli\'c, M.~Pu\v{s}i\'c, B.~Pokorni, Piecewise linear
  transformation in diffusive flux discretization, J. Comp. Phys. 282 (2015)
  227--237.

\bibitem{Yua08}
A.~Yuan, Z.~Sheng, Monotone finite volume schemes for diffusion equations on
  polygonal meshes, J. Comp. Phys. 227~(12) (2008) 6288--6312.

\bibitem{Dro14}
J.~Droniou, Finite volume schemes for diffusion equations: introduction to and
  review of modern methods, Math. Mod. Meth. Appl. Sci. 24~(8) (2014)
  1575--1619.

\bibitem{Mun10}
S.~S. Mundal, E.~Keilegavlen, I.~Aavatsmark, Simulation of anisotropic
  heterogeneous near-well flow using {MPFA} methods on flexible grids,
  Computat. Geosci. 14~(4) (2010) 509--525.

\bibitem{Che09}
Z.~Chen, Y.~Zhang, Well flow models for various numerical methods, Int. J.
  Numer. Anal. Mod. 6~(3) (2009) 375--388.

\bibitem{Din01}
Y.~Ding, L.~Jeannin, A new methodology for singular modeling in flow
  simulations in reservoir engineering, Computat. Geosci. 5~(2) (2001) 93--119.

\bibitem{Din04}
Y.~Ding, L.~Jeannin, New numerical schemes for near well modeling using
  flexible grids, SPE J. 9~(1) (2004) 109--121.

\bibitem{Dur00}
L.~J. Durlofsky, An approximate model for well productivity in heterogeneous
  porous media, Math. Geol. 32~(4) (2000) 421--438.

\bibitem{Pea78}
D.~Peaceman, Interpretation of well-block pressures in numerical reservoir
  simulation, SPE J. 18~(3) (1978) 183--194.

\bibitem{Pea83}
D.~Peaceman, Interpretation of wellblock pressures in numerical reservoir
  simulation with nonsquare grid blocks and anisotropic permeability, SPE J.
  23~(3) (1983) 531--543.

\bibitem{Hai95}
H.~M. Haitjema, Analytic Element Modeling of Groundwater Flow, Academic Press,
  Inc, San Diego, 1995.

\bibitem{Sut93}
S.~P. Sutera, R.~Skalak, The history of poiseuille's law, Annu. Rev. Fluid
  Mech. 25 (1993) 1--19.

\bibitem{Sib81}
R.~Sibson, A brief description of natural neighbour interpolation, in:
  V.~Barnet (Ed.), Interpreting multivariate data, Wiley, Chichester, 1981, pp.
  21--36.

\bibitem{Dot14}
M.~Dotli{\'c}, Finite volume methods for well-driven flows in anisotropic
  porous media, CMAM 14~(4) (2014) 473--483.

\bibitem{Dro11}
J.~Droniou, C.~{Le Potier}, Construction and convergence study of schemes
  preserving the elliptic local maximum principle, SIAM J. Numer. Anal. 49~(2)
  (2011) 459--490.

\bibitem{WODA}
D.~Vidovi\'c, M.~Dotli\'c, B.~Pokorni, {WODA solver},
  \url{http://www.sourceforge.net/projects/wodasolver/}.

\end{thebibliography}

\end{document}